\documentclass[12pt,letterpaper,english,preprint, aps,superscriptaddress]{revtex4}
\usepackage[T1]{fontenc}
\usepackage[latin9]{inputenc}
\usepackage[dvipdf]{color}
\usepackage{amsmath}
\usepackage{amssymb}
\usepackage{epstopdf}
\usepackage{graphicx}



\usepackage{babel}
\begin{document}

\title{The projections on $ZZ\gamma$ and $Z\gamma\gamma$ couplings via $\nu\bar \nu \gamma$ production in HL-LHC and HE-LHC
}

\author{A. Senol}
\email{senol_a@ ibu.edu.tr}
\affiliation{Department of Physics, Bolu Abant Izzet Baysal University, 14280, Bolu, Turkey}
\author{H. Denizli}
\email{denizli_h@ ibu.edu.tr}
\affiliation{Department of Physics, Bolu Abant Izzet Baysal University, 14280, Bolu, Turkey}

\author{A. Yilmaz}
\email{aliyilmaz@giresun.edu.tr}

\affiliation{Department of Electrical and Electronics Engineering, Giresun University, 28200 Giresun, Turkey}

\author{I. Turk Cakir}
\email{ilkay.turk.cakir@cern.ch}
\affiliation{Department of Energy Systems Engineering, Giresun University, 28200 Giresun, Turkey}

\author{O. Cakir}
\email{ocakir@science.ankara.edu.tr}
\affiliation{Department of Physics, Ankara University, 06100 Ankara, Turkey.}
\date{\today}\date{\today}
\begin{abstract}
We investigate the sensitivity of the anomalous dimension-8 neutral triple gauge couplings via process $pp\to \nu\nu\gamma$ with fast detector simulation including pile-up effects for the post LHC experiments. The transverse momentum of the final state photon and missing energy transverse distributions are considered in the analysis.  We obtain the sensitivity to the $C_{\widetilde B W}/\Lambda^4$, $C_{B B}/\Lambda^4$, $C_{WW}/\Lambda^4$  and $C_{BW}/\Lambda^4$ couplings at HL-LHC and HE-LHC with an integrated luminosity of 3 ab$^{-1}$ and 15 ab$^{-1}$, respectively. Finally, our numerical results show that one can reach the constraints at 95\% confidence level without systematic error on $C_{\widetilde BW}/\Lambda^4$, $C_{B B}/\Lambda^4$, $C_{W W}/\Lambda^4$  and $C_{BW}/\Lambda^4$ couplings for HL-LHC (HE-LHC) as [-0.38;0.38] ([-0.12;0.12]), [-0.21;0.21]([-0.085;0.085]), [-1.08;1.08]([-0.38;0.38]) and [-0.48;0.48]([-0.25;0.25]), respectively. They are better than the experimental limits obtained by LHC .

\end{abstract}

\maketitle

\section{Introduction}


The Standard Model (SM) is a successful theory at describing the particle physics phenomena in reachable energy limits of current collider experiments. Nevertheless, the SM needs to be extended to explain some experimental and theoretical facts such as the electroweak scale-Planck scale hierarchy problem, the striking evidence of dark matter, the dynamic origin of the Higgs mechanism and non-zero neutrino masses. Any significant deviation from the expectation of SM has not been observed since the beginning of the LHC era. However, the explanation of physics beyond the SM can benefit from the High Luminosity LHC (HL-LHC) \cite{Apollinari:2015bam} and High Energy LHC (HE-LHC) \cite{Abada:2019ono} with novel technologies and approaches. These future considerations aim to extend direct and indirect sensitivities to new physics and discoveries not only by increasing luminosity and center of mass energy but also sensitive technologies involved in the detectors. 

The progress in the understanding of the Higgs and gauge sector via the precision measurements can play key role in exploration of the new physics.  A new range of luminosity and energy will provide valuable information to find new phenomena or set limits on various aspects of new physics models in coming years. In this work, we use well motivated effective field theory (EFT) to probe expected deviations on the neutral three-boson couplings from new physics.

 We use a Lagrangian of the EFT including both SM interactions and neutral Triple
Gauge Couplings (NTGC) obeying local $U(1)_{EM}$ and Lorentz symmetry as given in Ref. \cite{Degrande:2013kka}

\begin{eqnarray}
\mathcal{L}^{nTGC}=\mathcal{L}^{SM}+\sum_i\frac{C_i}{\Lambda^4}(\mathcal{O}_i+\mathcal{O}_i^{\dagger})
\end{eqnarray}
where four operators with index $i$ can be expressed as
\begin{eqnarray}
\mathcal{O}_{BW}&=&iH^{\dagger}B_{\mu\nu}W^{\mu\rho}\{D_{\rho},D^{\nu}\}H\\
\mathcal{O}_{WW}&=&iH^{\dagger}W_{\mu\nu}W^{\mu\rho}\{D_{\rho},D^{\nu}\}H\\
\mathcal{O}_{BB}&=&iH^{\dagger}B_{\mu\nu}B^{\mu\rho}\{D_{\rho},D^{\nu}\}H\\
\mathcal{O}_{\tilde B W}&=&iH^{\dagger} \tilde B_{\mu\nu}W^{\mu\rho}\{D_{\rho},D^{\nu}\}H
\end{eqnarray}
where $\tilde B_{\mu\nu}$ is a dual $B$ strength tensor. The following convention in the definitions of the operators are used: 
\begin{eqnarray}
W_{\mu\nu}&=&\sigma^I(\partial_{\mu} W_{\nu}^I-\partial_{\nu} W_{\mu}^I+g\epsilon_{IJK}W_{\mu}^JW_{\nu}^K)\\
B_{\mu\nu}&=&(\partial_{\mu} B_{\nu}-\partial_{\nu} B_{\mu})
\end{eqnarray}
with $\left<\sigma^I\sigma^J\right>=\delta^{IJ}/2$
and 
\begin{eqnarray}
D_{\mu}\equiv\partial_{\mu}-ig_wW_{\mu}^i\sigma^i-i\frac{g'}{2}B_{\mu}Y
\end{eqnarray}
These dimension-eight operators have four coefficients. The $C_{\tilde{B}W} / \Lambda^{4}$coefficient is CP-conserving, while the others $C_{BB}/ \Lambda^{4}$, $C_{BW}/ \Lambda^{4}$, $C_{WW}/ \Lambda^{4}$ are CP-violating. As described in Ref. \cite{Degrande:2013kka}., they can also be related to dimension-six operators of anomalous NTGC (aNTGC).
The dimension-six operators can have an effect on aNTGC at one-loop level (at the order  $O(\alpha \hat s /4\pi \Lambda^2)$) as they do not induce aNTGC at tree-level \cite{Degrande:2013kka}. However, the tree level contributions from dimension-eight operators are of the order $O(\hat s ν^2/\Lambda^4)$. Therefore, one-loop contribution of the dim-6 operators can be ignored with respect to that of dim-8 operators for $\Lambda < 2v\sqrt{\pi/\alpha}$.

The ATLAS collaboration sets the current experimental bounds on dimension-eight operators with a conversion from the coefficients of dimension-six operators for the process $pp\to ZZ\to l^+l^-l'^+l'^-$ \cite{Aaboud:2017rwm}. The production process of $pp\to Z\gamma \to\nu\bar \nu$ has already been searched in Ref. \cite{ATLAS:2018eke} at $\sqrt s$ = 13 TeV with $L_{int}$=36.1 fb$^{-1}$. The results from these references are tabulated at a 95\% C.L. in Table \ref{tab1}.

\begin{table}
\caption{The current experimental limits on Dimension-8 aNTGC from ATLAS collaborations at the 95\% C.L. \label{tab1}}
\begin{ruledtabular}
\begin{tabular}{lcccc}
 Couplings (TeV$^{-4}$)   & &  $ZZ \rightarrow 4 \ell$  \cite{Aaboud:2017rwm} & $Z\gamma \rightarrow \nu \bar{\nu}\gamma$~\cite{ATLAS:2018eke} \\ \hline
$C_{\tilde{B}W}  / \Lambda^{4}$ & & $-5.9, +5.9$ & $-1.1, \,\, +1.1$ \\
$C_{WW} / \Lambda^{4}$ & & $-3.0, +3.0$ & $-2.3, \,\, +2.3$ \\
$C_{BW} / \Lambda^{4}$ & & $-3.3, +3.3$ &  $-0.65, +0.64$\\
$C_{BB} / \Lambda^{4}$ & & $-2.7, +2.8$ & $-0.24, +0.24$ \\
\end{tabular}
\end{ruledtabular}
\end{table}

The physics programme accessible at the HL-LHC and HE-LHC extends beyond the LHC and covers pp collisions at 14 TeV and 27 TeV with an integrated luminosity of 3 ab$^{-1}$ and 15 ab$^{-1}$ each for the LHC experiments, respectively.


The constraints on dimension eight operatos in the $pp\to ll\gamma$ and $pp\to \nu\nu\gamma$ processes have been studied for a 100 TeV center of mass energy collider (FCC-hh) \cite{Senol:2018cks}. In this analysis we focus on CP-conserving $C_{\widetilde BW}/\Lambda^4$ and CP-violating $C_{BB}/\Lambda^4$, $C_{WW}/\Lambda^4$,$C_{BW}/\Lambda^4$ couplings via $pp\to \nu\nu\gamma$ process in other post LHC considerations, namely HL-LHC and HE-LHC including pile-up effects.
$Z(\nu\bar {\nu})\gamma$ final state is selected due to its several advantages over the $Z(ll)\gamma$ and $Z(q\bar q)\gamma$ final states. Even though $Z$ decaying to quark-antiquark pair has a branching ratio of 69.9\%, this channel is contaminated by a large multi-jet background. $Z$ boson decays in 20\% of the cases into neutrino pair while 10\% into charged lepton-antilepton pairs. Therefore one can take advantages of the opportunity to study $Z\gamma$ production in more energetic (higher $E_T^{\gamma}$) region. In addition,  this process is more sensitive to bosonic couplings.

\begin{figure}[htb!]
\includegraphics{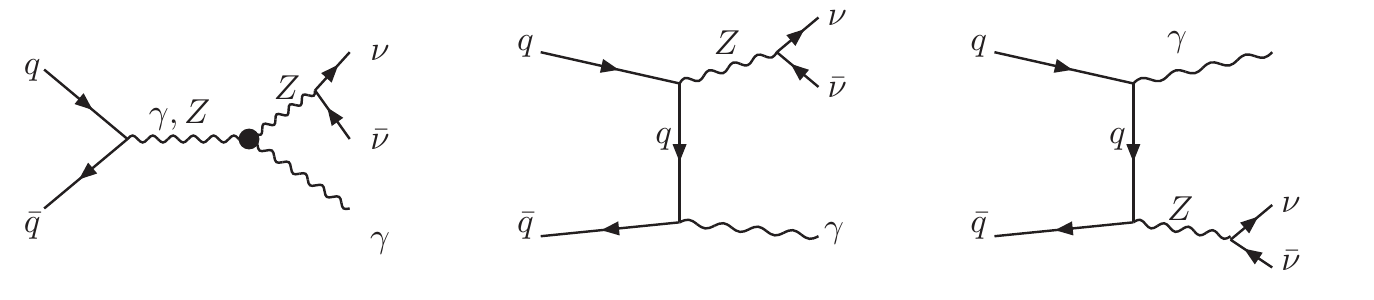}
\caption{ Feynman diagrams for $pp\to \nu\bar \nu\gamma$ process;  the anomalous contribution of $ZZ\gamma$ and $Z\gamma\gamma$ vertices (first diagram) and the SM contribution.  \label{fd_2}}
\end{figure}

\begin{figure}[htb!]
\includegraphics[scale=0.64]{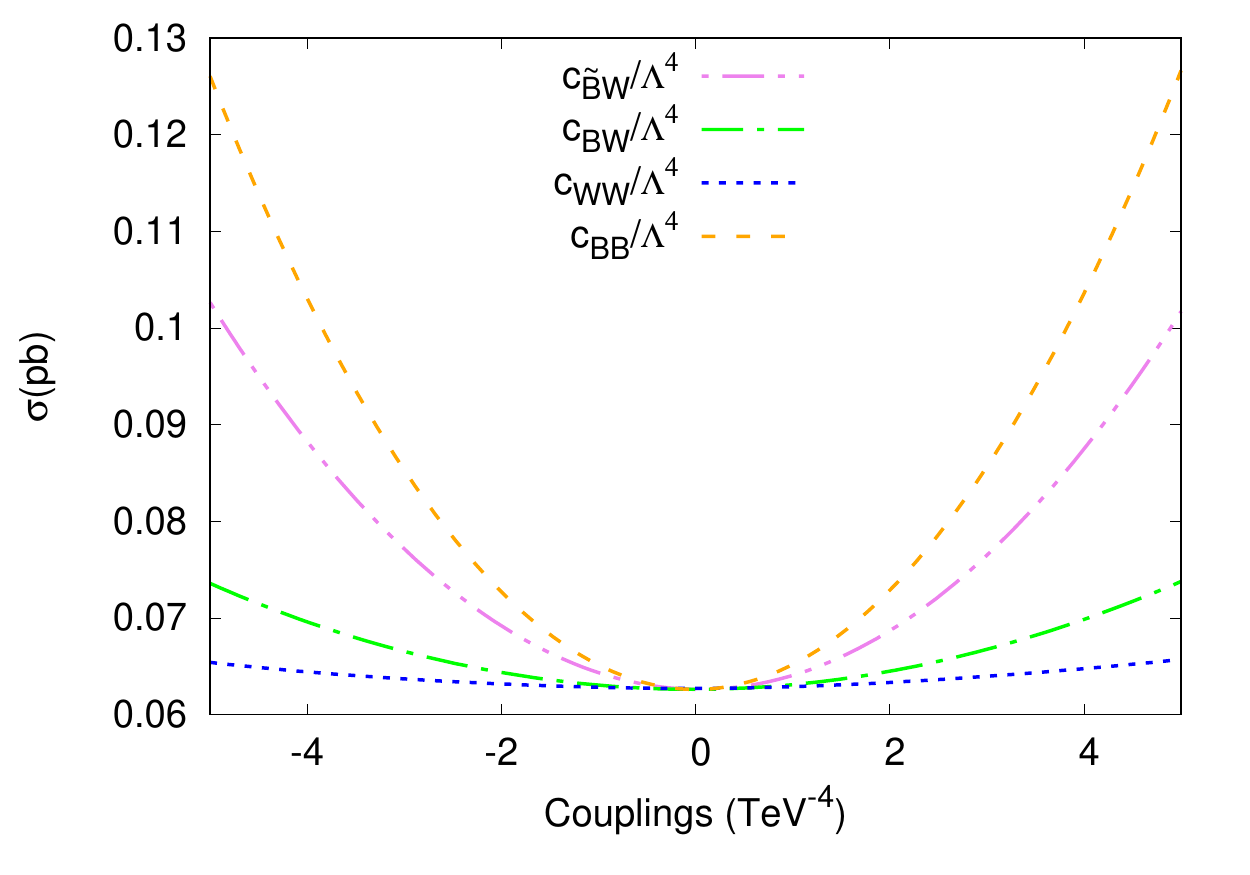} 
\includegraphics[scale=0.64]{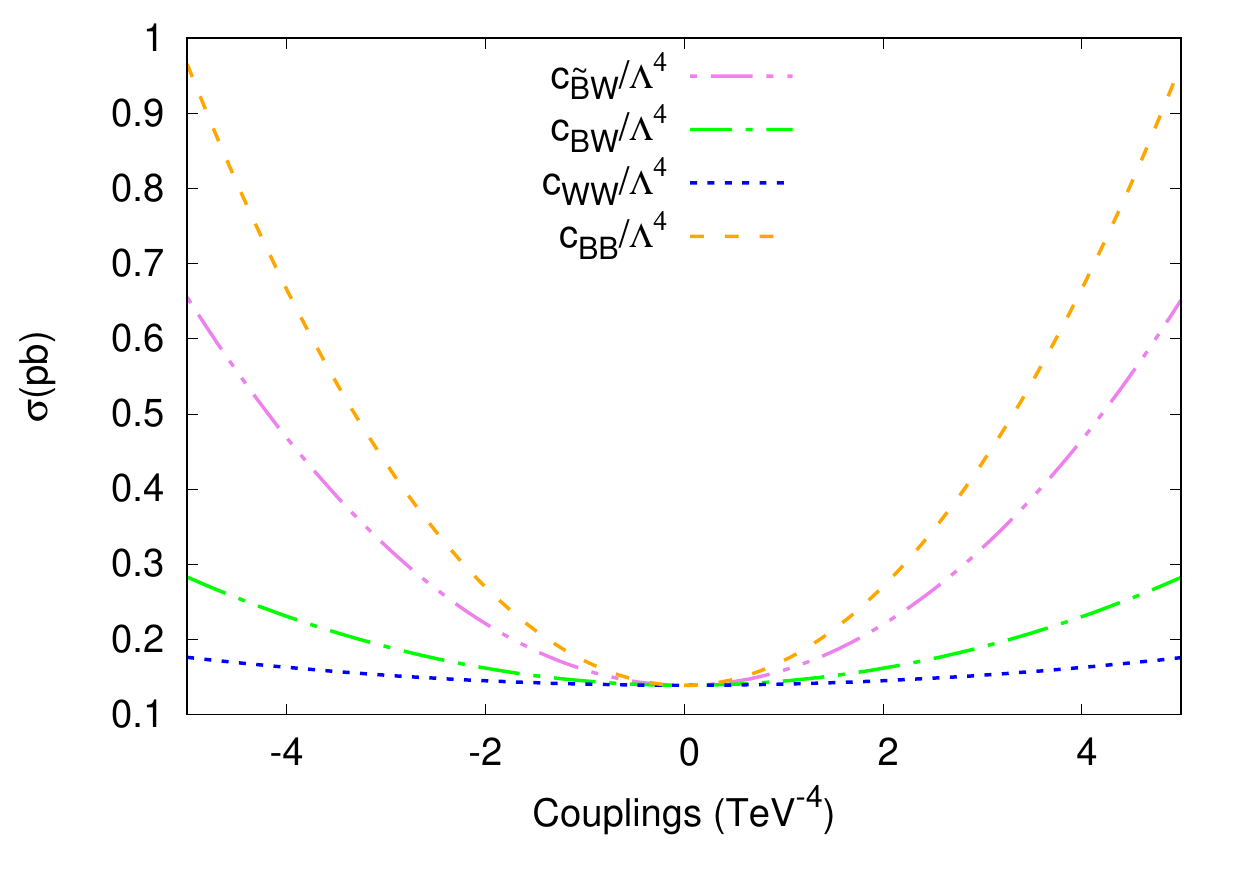} 
\caption{ The signal cross sections of $pp\to \nu\bar \nu\gamma$ depending on dimension-eight couplings
at HL-LHC (on the left) and HE-LHC (on the right).  \label{crosssection} }
\end{figure}

The tree level Feynman diagrams of the $pp\to \nu\nu\gamma$ process are shown in Fig.~\ref{fd_2}. 
The first  diagram of this figure accounts for the anomalous $Z\gamma\gamma$ and $ZZ\gamma$ couplings, while the others for SM contributions. The sensitivities are investigated using the Monte Carlo simulations with a leading order in \verb|MadGraph5_aMC@NLO| \cite{Alwall:2014hca}. The operators described in Eqs.(2)-(5) are implemented into \verb|MadGraph5_aMC@NLO| through Feynrules package \cite{Alloul:2013bka}  as a Universal FeynRules Output (UFO) module \cite{Degrande:2011ua}. The cross sections of the $pp\to \nu\nu\gamma$ process as a function of  $C_{\widetilde BW}/\Lambda^4$,$C_{BB}/\Lambda^4$, $C_{WW}/\Lambda^4$ and $C_{BW}/\Lambda^4$ couplings for HL-LHC on the left panel and HE-LHC on the right panel are shown in Fig.2. In this figure, the cross sections are calculated at leading order including the transverse momentum ( $p_T^{\gamma}$ >100 GeV) and pseudo-rapidity ($\eta^{\gamma}$< 2.5) cuts for photons. In addition, one of the effective couplings is non-zero at a time, while the other couplings are fixed to zero. As it can be seen from Fig.~\ref{crosssection},  deviation from SM value of the anomalous cross section including    $C_{\widetilde BW}/\Lambda^4$,$C_{BB}/\Lambda^4$ couplings is larger than that for $C_{WW}/\Lambda^4$  and $C_{BW}/\Lambda^4$ in both HL-LHC and HE-LHC options. The cross sections values for the HE-LHC option are greater than for the HL-LHC option as one can expect due to higher center of mass energy. We will consider $C_{\widetilde B W}$, $C_{BB}$, $C_{BW}$ and $C_{WW}$ couplings in the detailed analysis including detector and pile-up effects through $\nu\nu\gamma$ production at HL-LHC and HE-LHC with 14 TeV  and 27 TeV center of mass energy in the next section.


\section{Signal and Background Analysis }
The detailed analysis of effective dim-8  aNTGC couplings and SM contributions as well as interference between effective couplings and SM contribution is performed via $pp\to \nu\nu\gamma$ process. Approximately two million events are  generated at LO partonic level in \verb|MadGraph5_aMC@NLO|  applying pseudo-rapidity ($|\eta^{\gamma}|$< 2.5) and transverse momentum ( $p_T^{\gamma}$ >20 GeV) cuts for both SM background as well as signal different with values of $C_{\widetilde BW}$, $C_{BB}$, $C_{BW}$ and $C_{WW}$  couplings. These events are passed through Pythia 8 \cite{Sjostrand:2014zea} including initial and final parton shower and the fragmentation. Since pile-up events are major issue at the HL-LHC and HE-LHC, the detector responses are taken into account with upgrade card, namely \verb|CMS_phaseII_140PU_conf4.tcl| within Delphes \cite{deFavereau:2013fsa} package.  This card includes an average number, $\mu$, of 140 proton-proton interactions in the same bunch crossing (pileup) which corresponds to the nominal luminosity for HL-LHC and HE-LHC.  All events are analysed by using the ExRootAnalysis utility with ROOT \cite{Brun:1997pa}. The kinematical distributions are normalized to the number of expected events (the cross section of each processes times integrated luminosity of $L_{int}$=3 ab$^{-1}$ and $L_{int}$=15 ab$^{-1}$  for HL-LHC and HE-LHC, respectively).

\begin{figure}[htb!]
\includegraphics[scale=0.2]{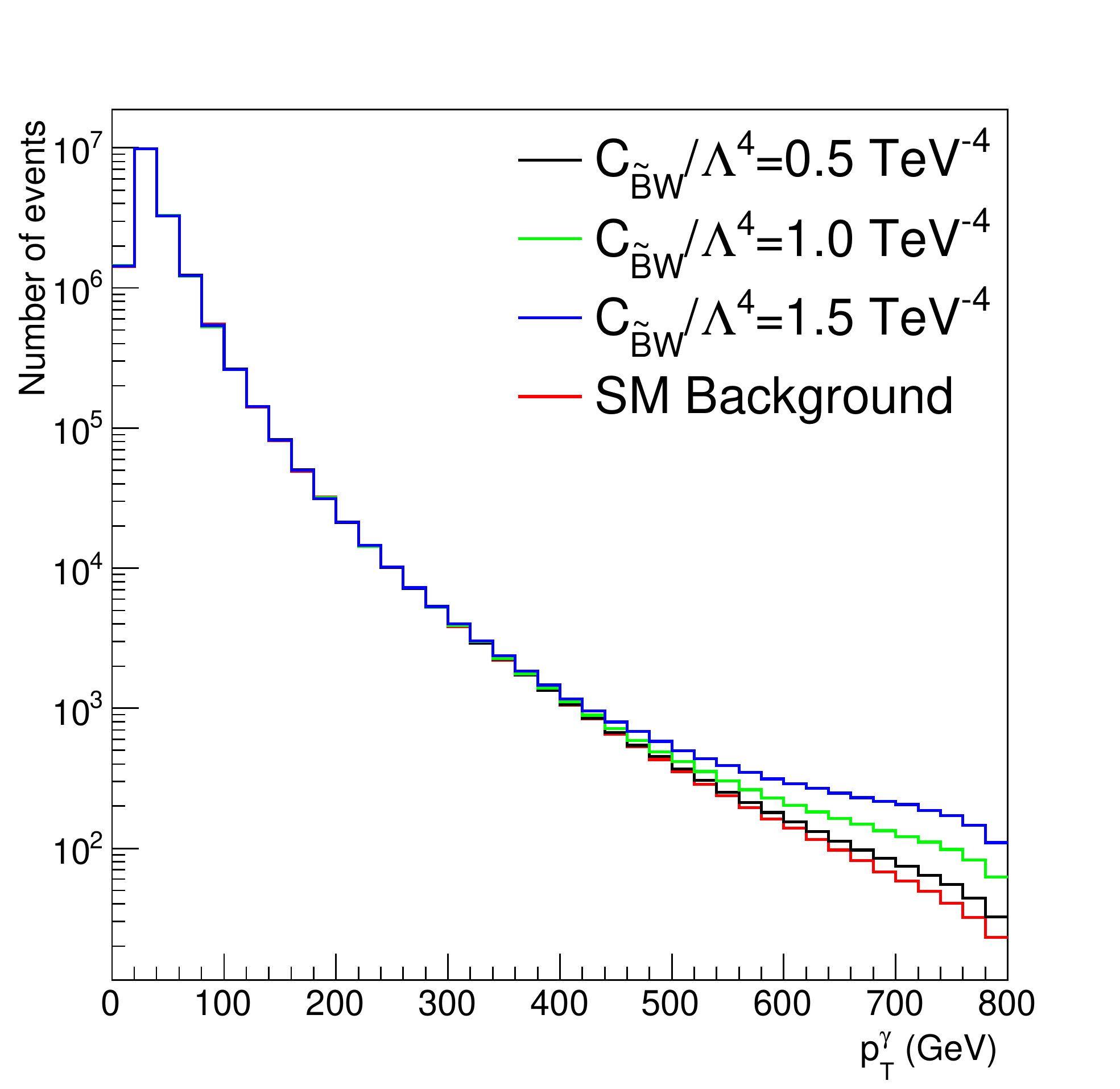} 
\includegraphics[scale=0.2]{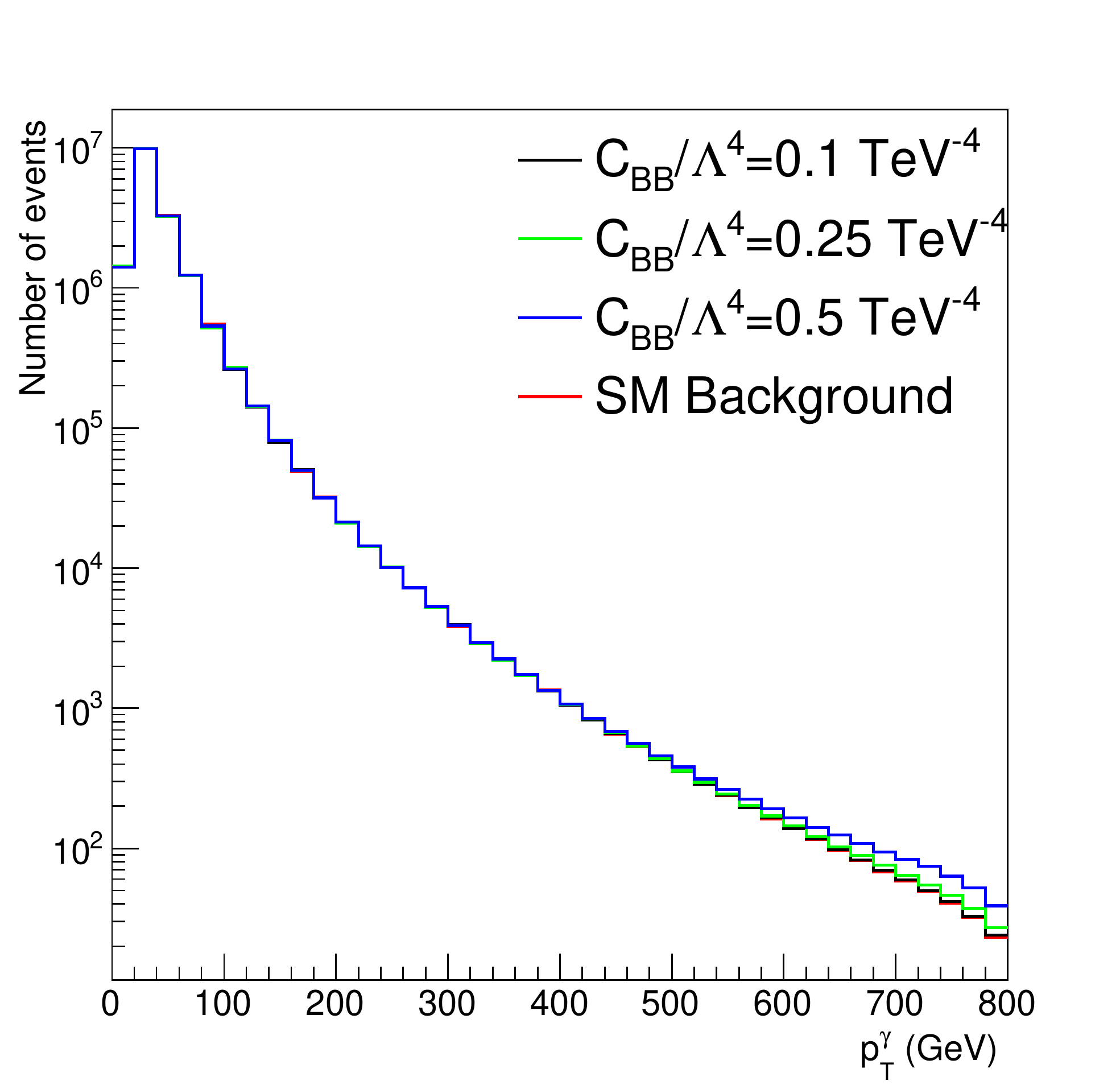} 
\includegraphics[scale=0.2]{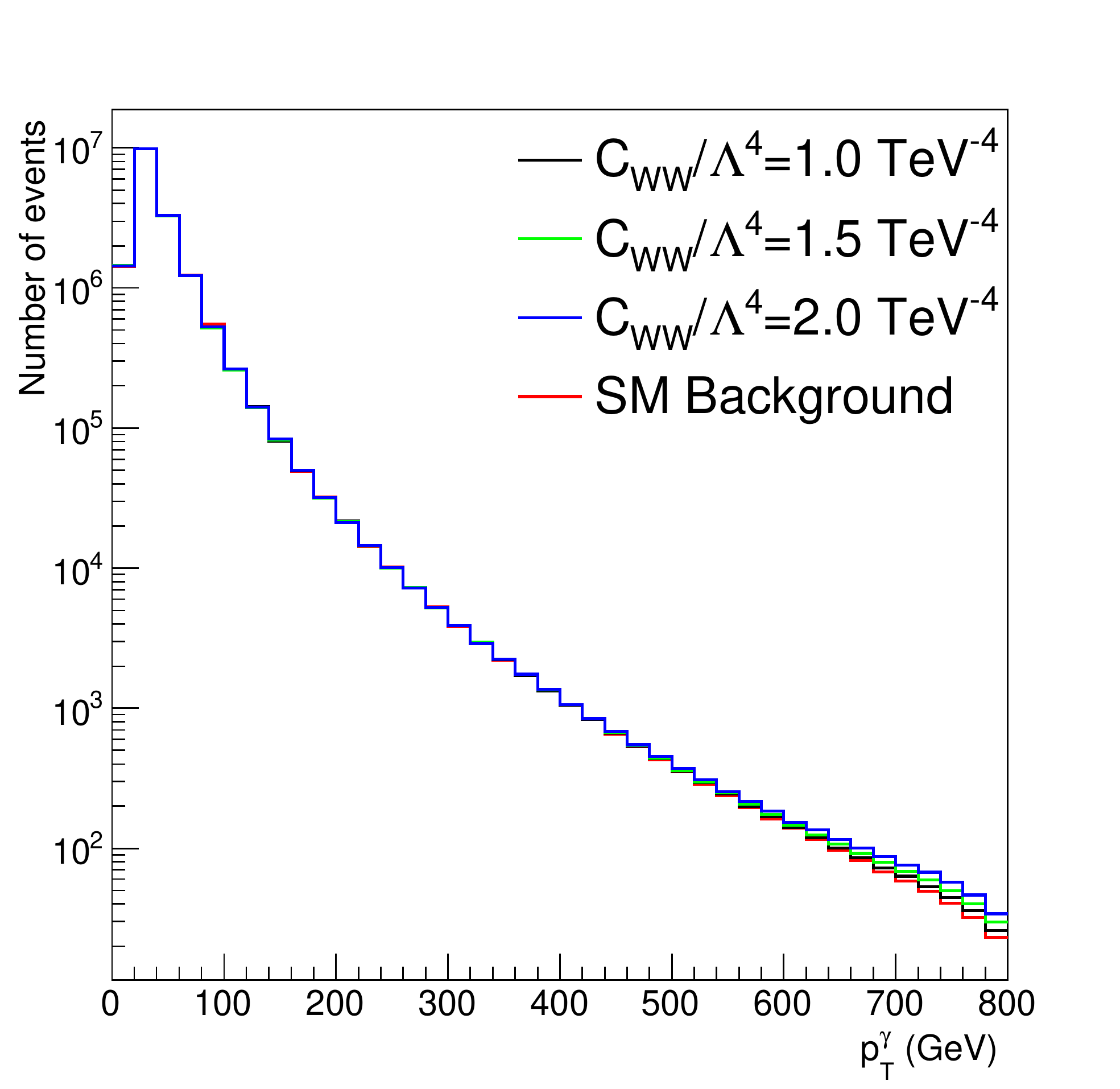} 
\includegraphics[scale=0.2]{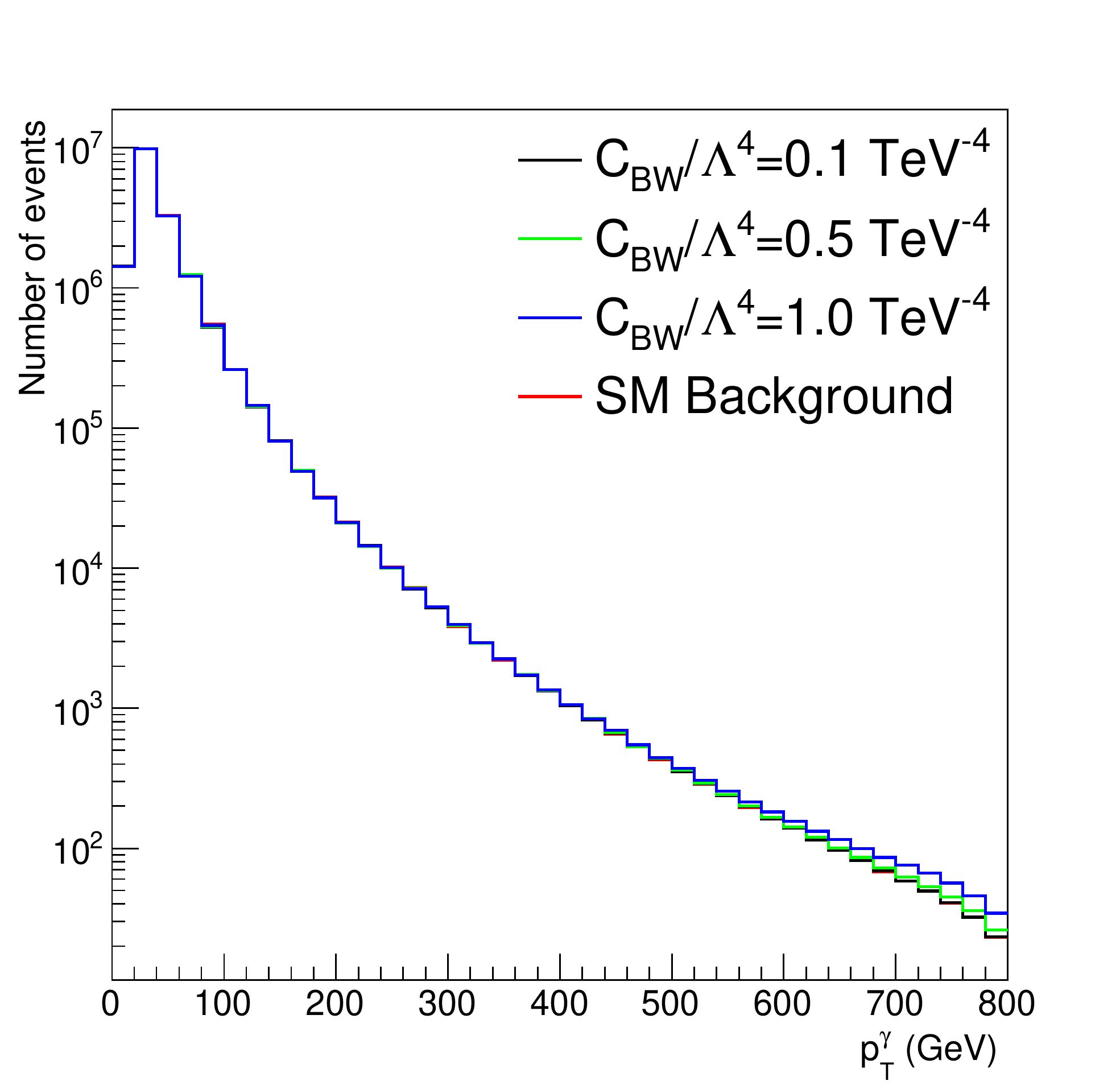} \\
\includegraphics[scale=0.2]{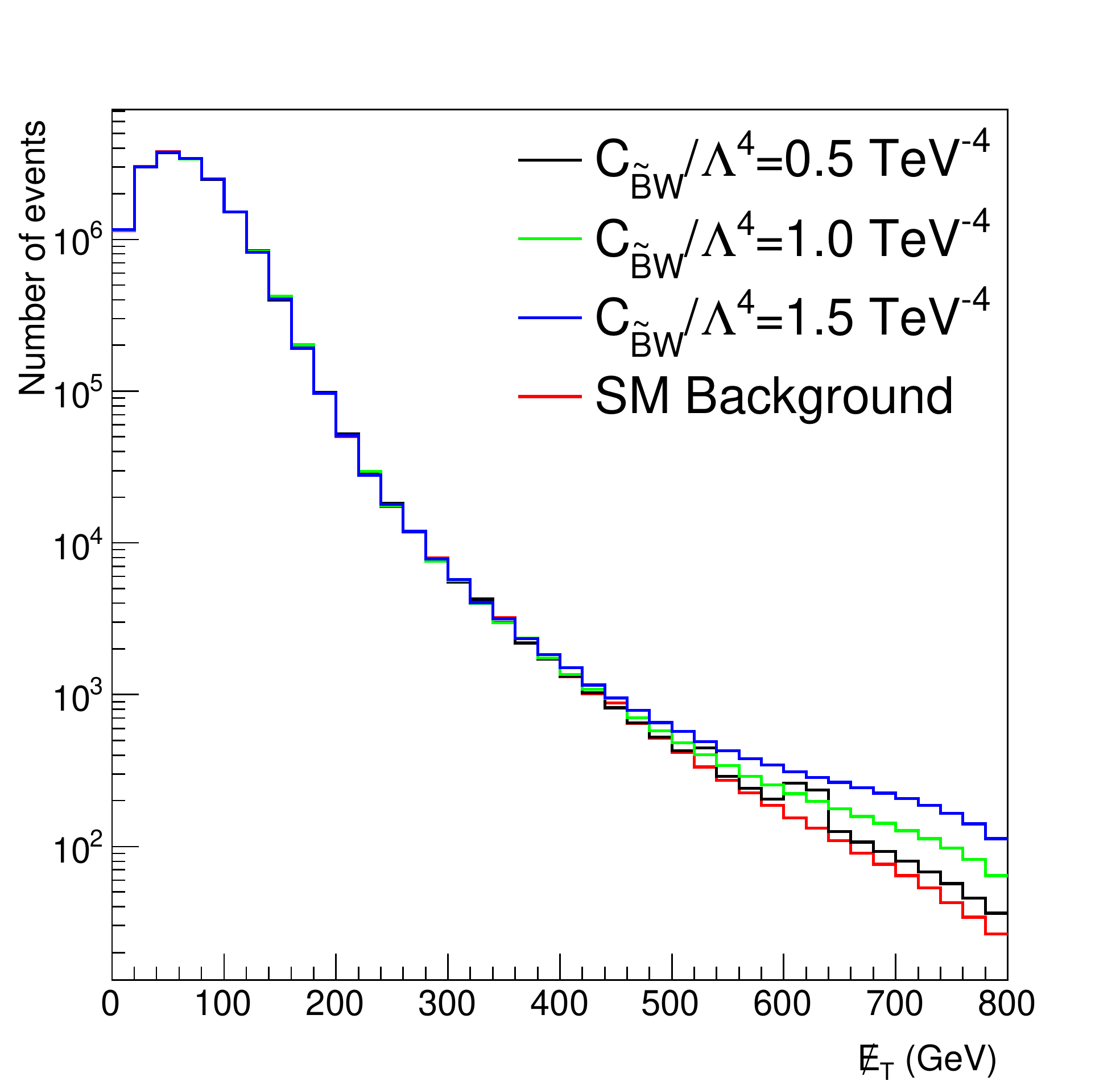} 
\includegraphics[scale=0.2]{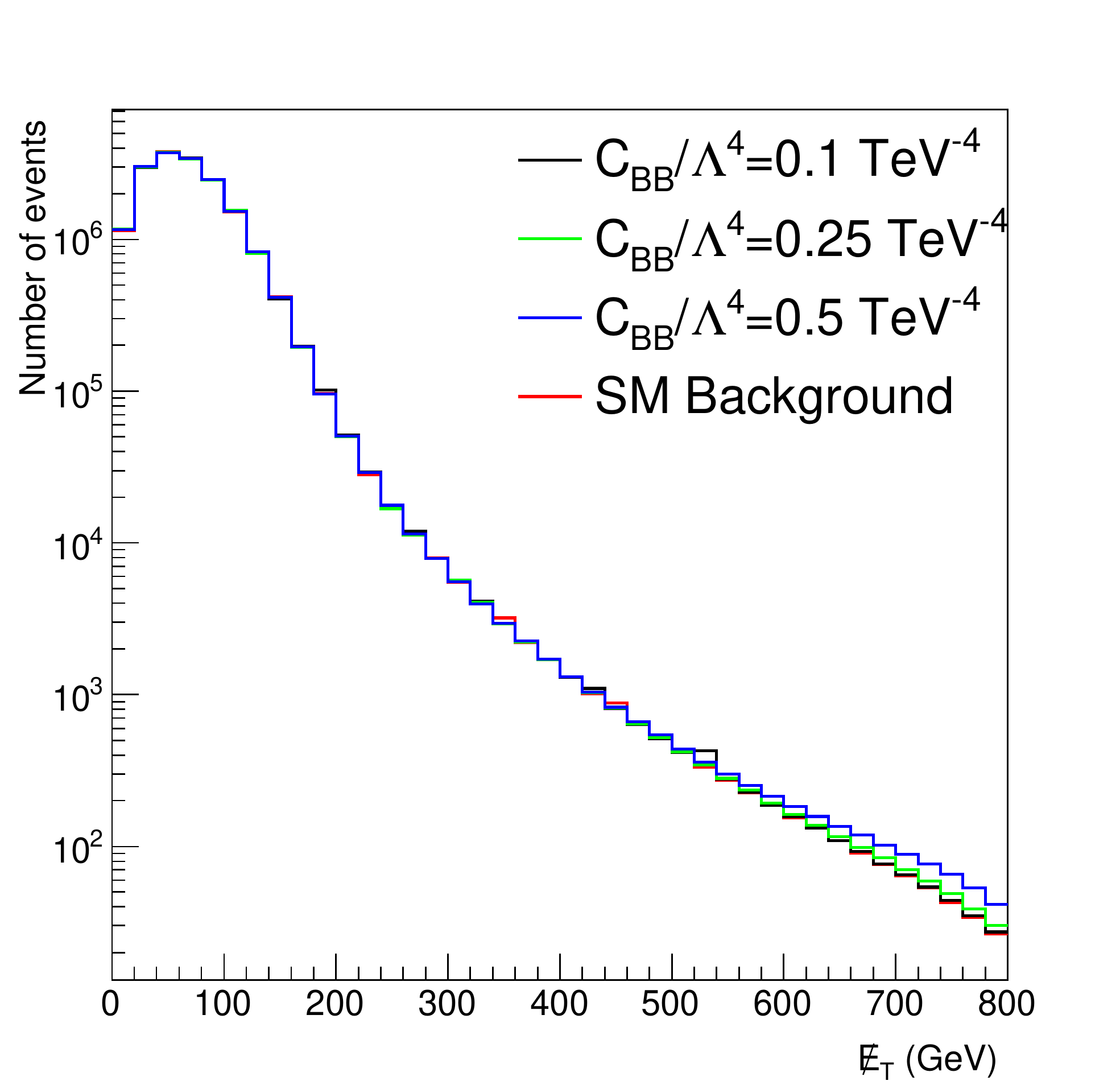} 
\includegraphics[scale=0.2]{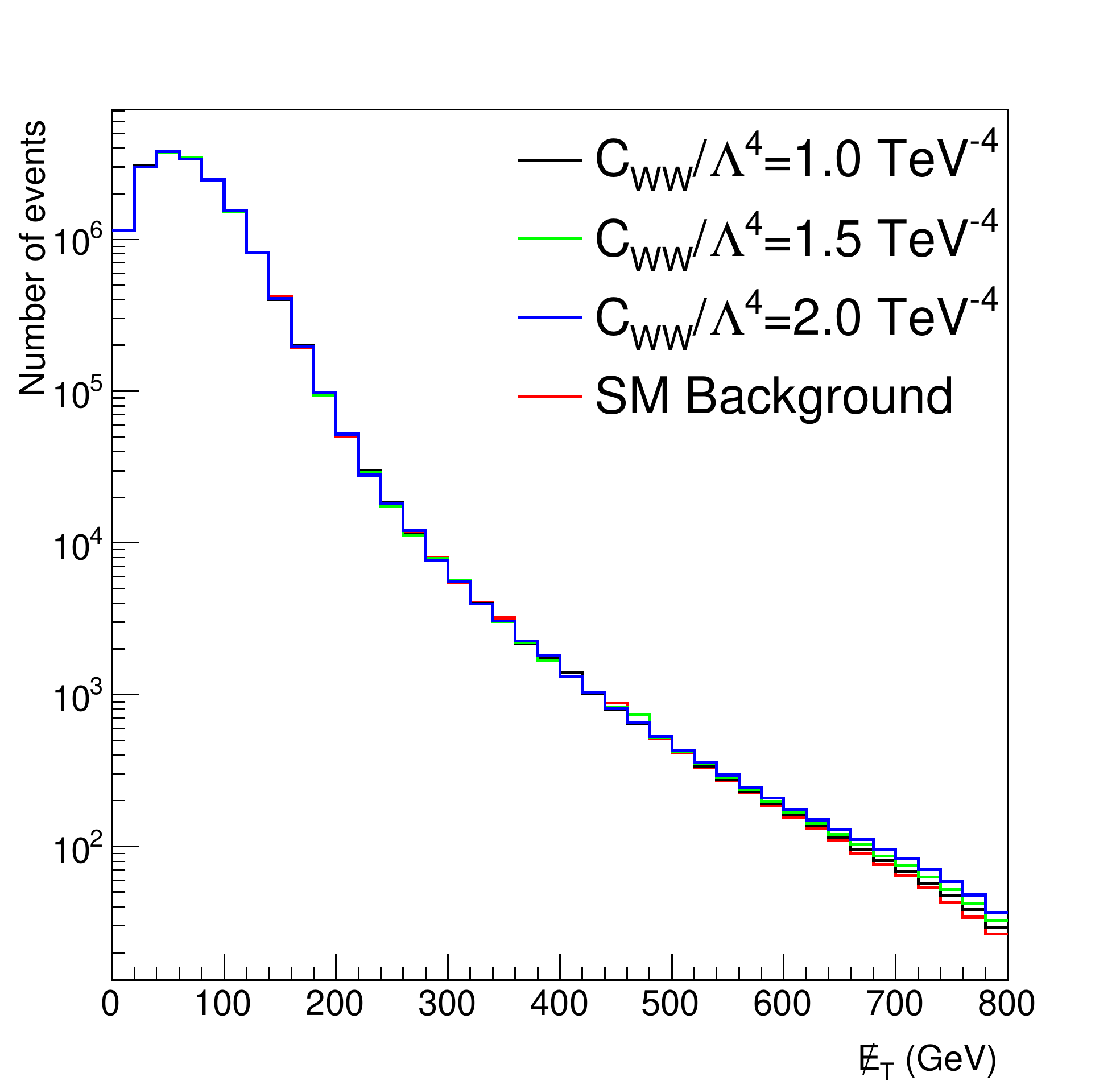} 
\includegraphics[scale=0.2]{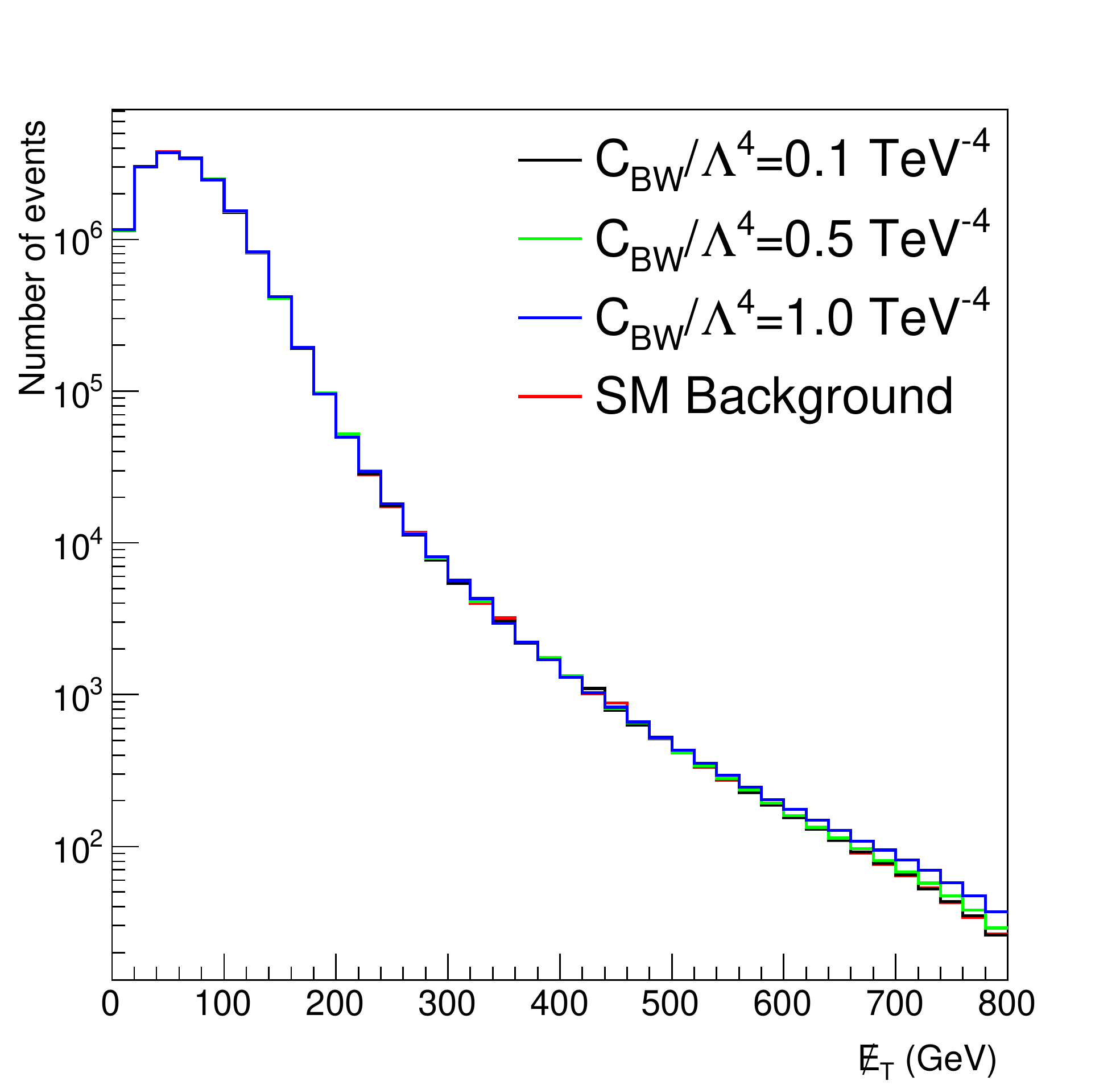} 
\caption{The normalized $p_T^{\gamma}$ distribution of leading photon (upper) and MET distribution (lower) for signal $C_{\widetilde BW}/\Lambda^4$,$C_{BB}/\Lambda^4$, $C_{WW}/\Lambda^4$  and $C_{BW}/\Lambda^4$ (left to right) couplings  and corresponding SM background for $pp\to \nu\bar \nu \gamma$ process at HL-LHC.  \label{ptmet_HLLHC}}
\end{figure}

\begin{figure}[htb!]
\includegraphics[scale=0.2]{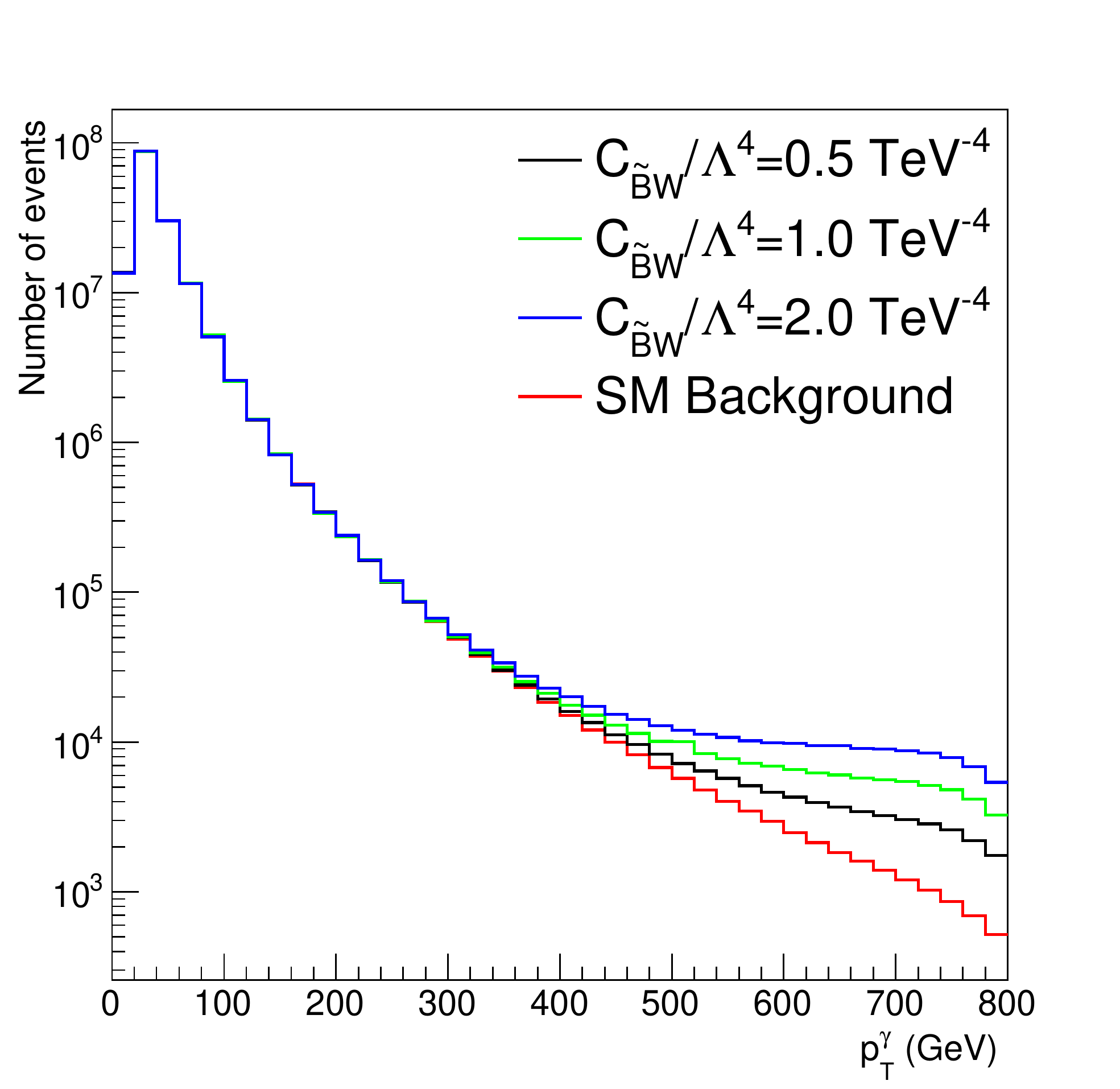} 
\includegraphics[scale=0.2]{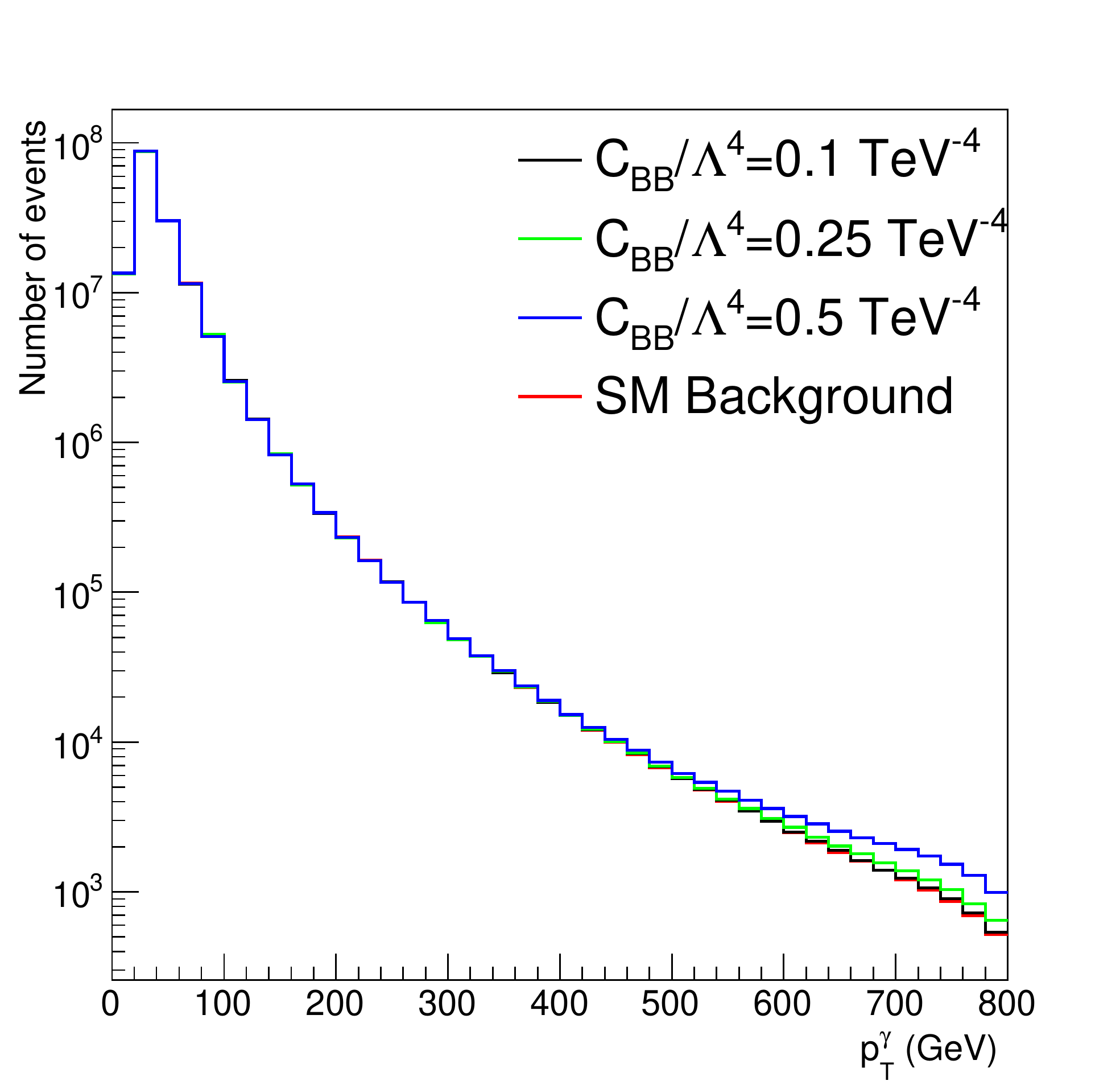} 
\includegraphics[scale=0.2]{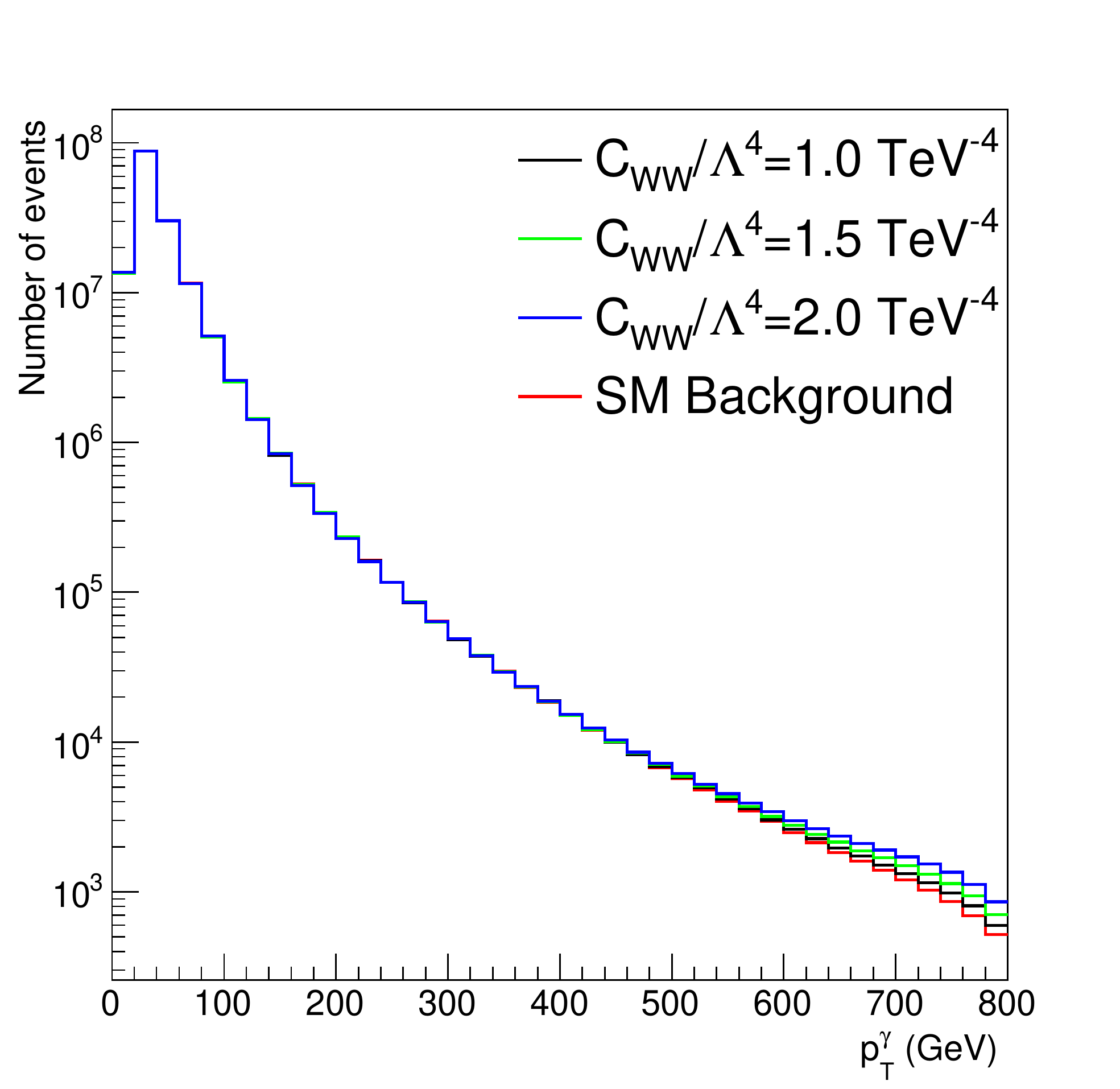} 
\includegraphics[scale=0.2]{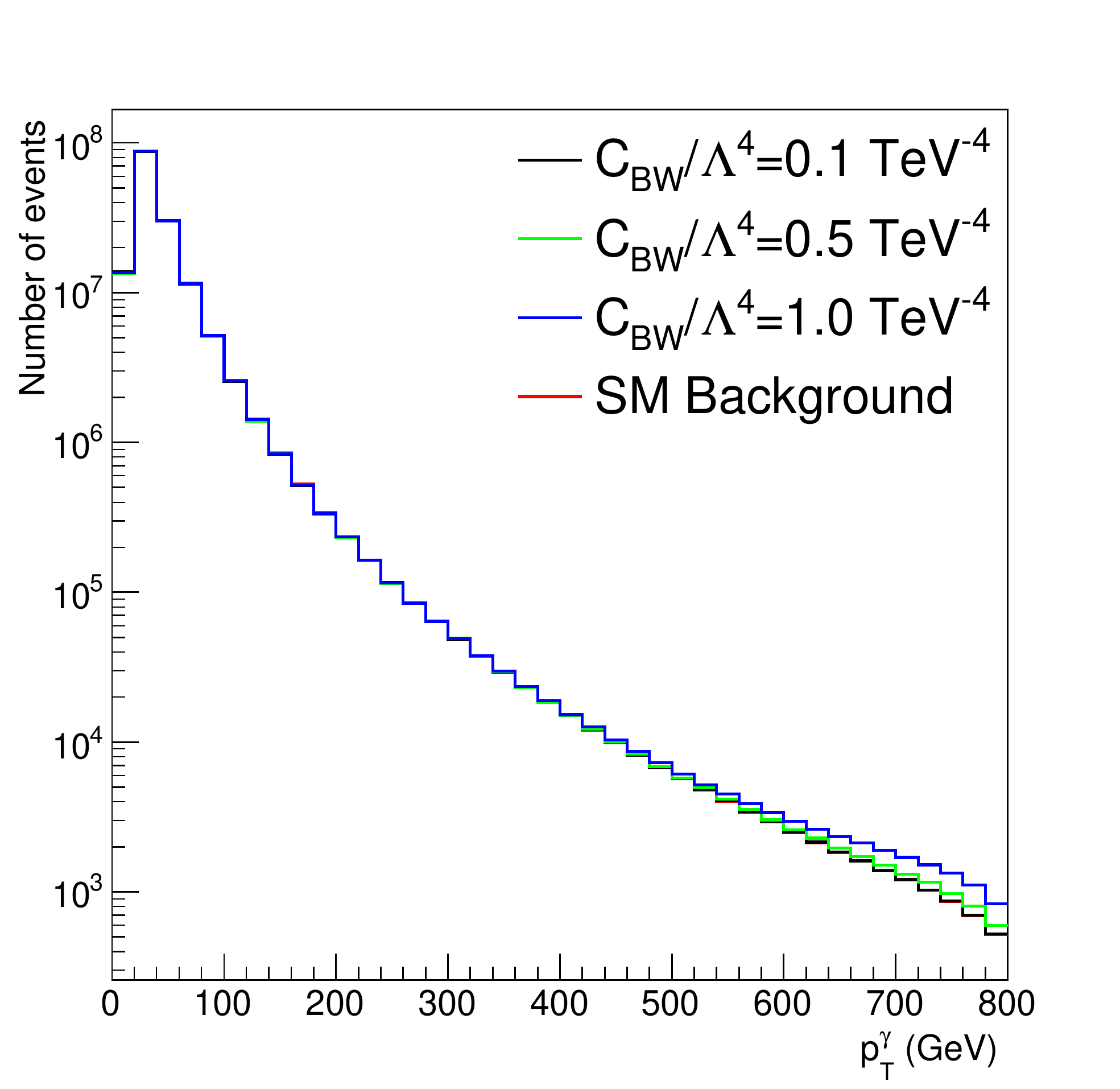} \\
\includegraphics[scale=0.2]{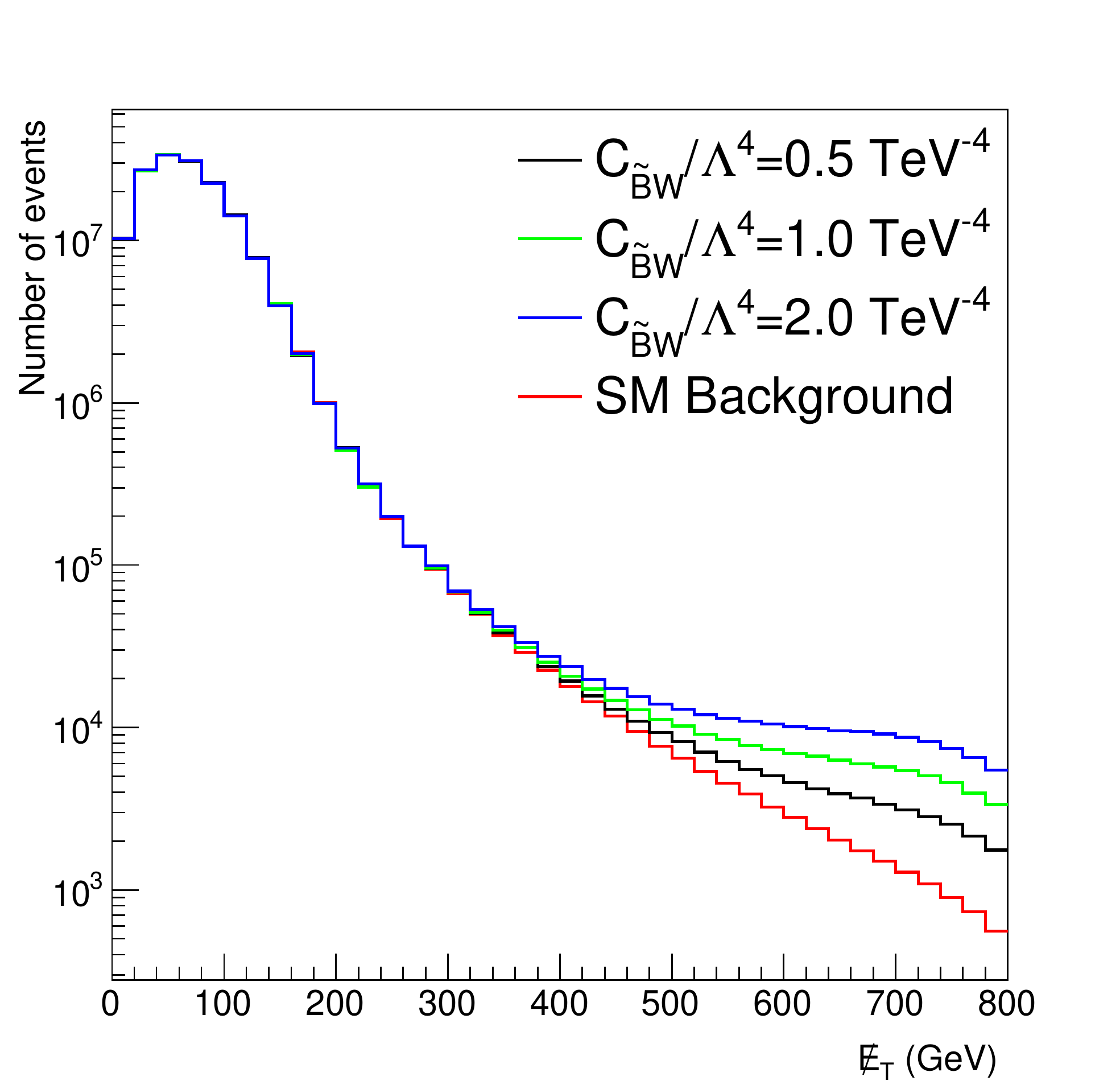} 
\includegraphics[scale=0.2]{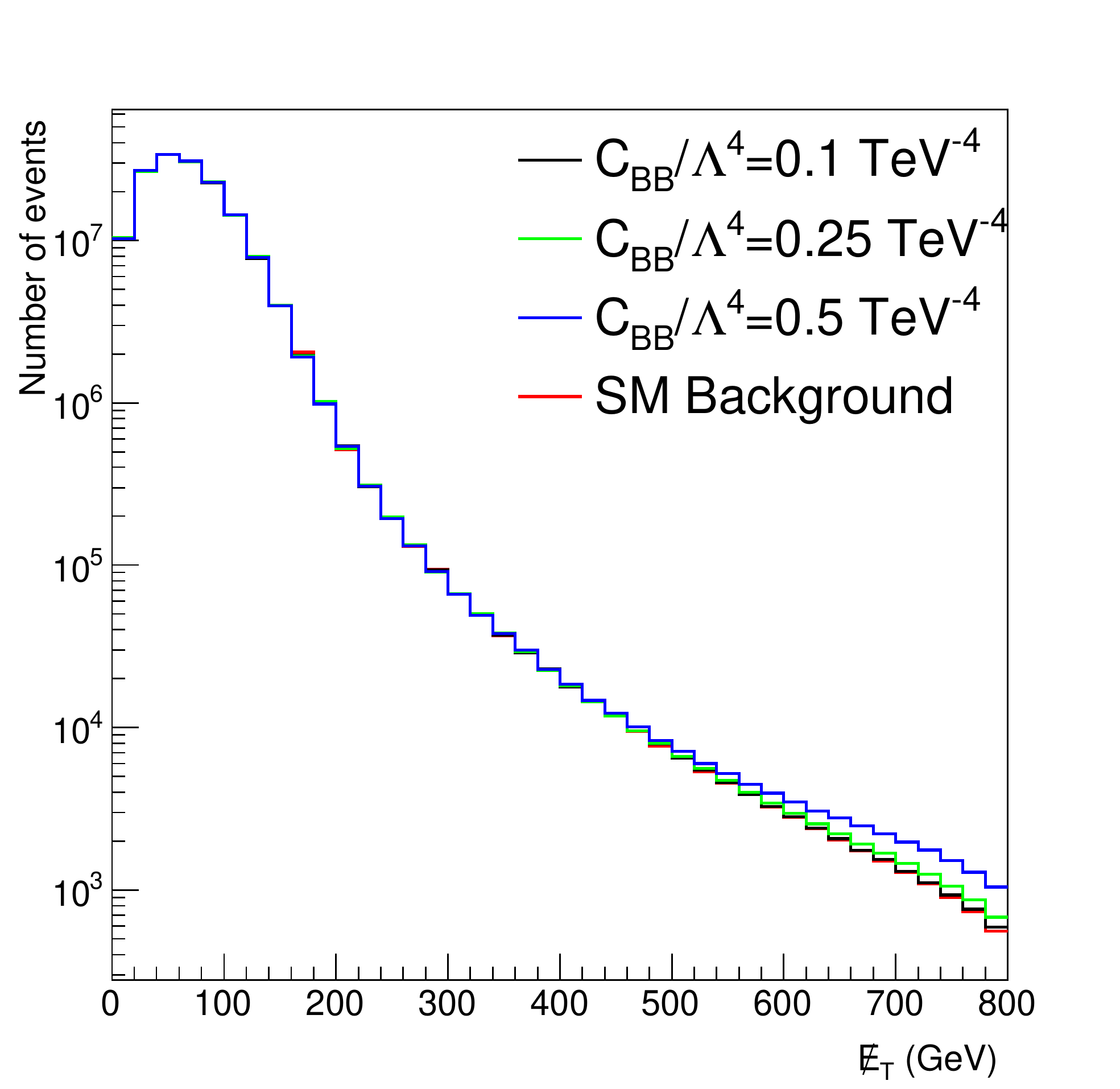} 
\includegraphics[scale=0.2]{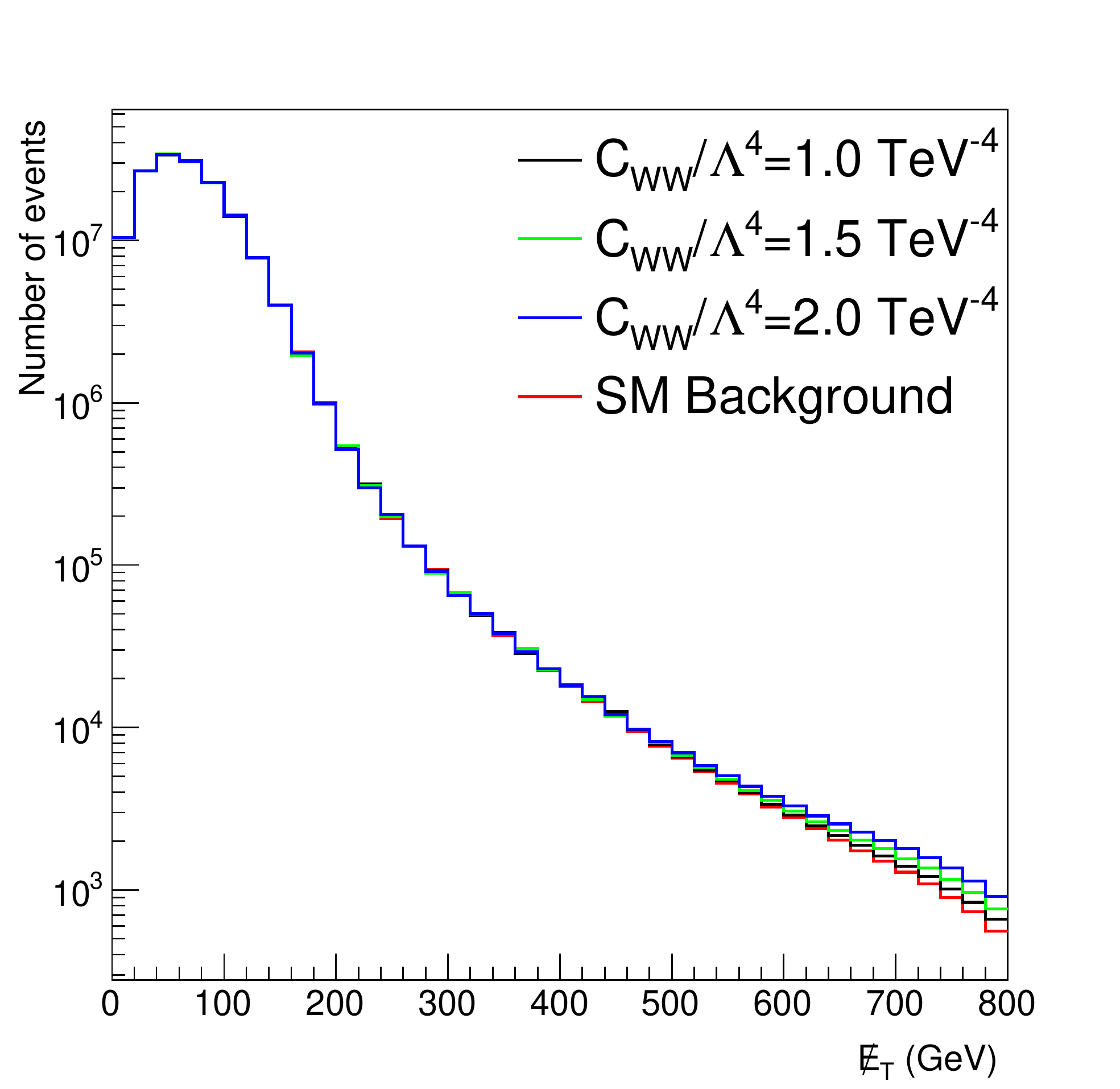} 
\includegraphics[scale=0.2]{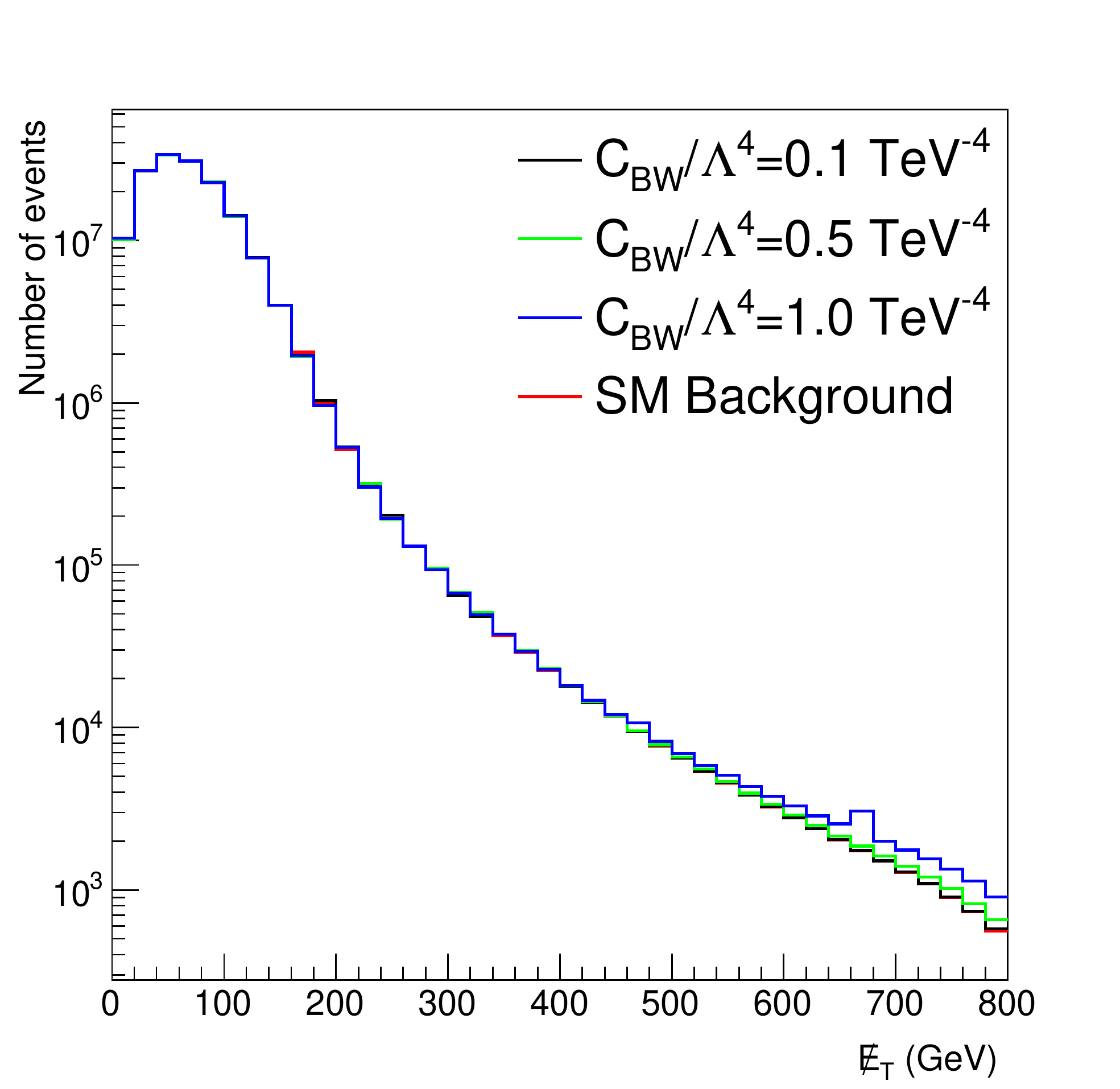} 
\caption{The normalized $p_T^{\gamma}$ distribution of leading photon (upper) and MET distribution (lower) for signal  $C_{\widetilde BW}/\Lambda^4$,$C_{BB}/\Lambda^4$, $C_{WW}/\Lambda^4$  and $C_{BW}/\Lambda^4$ (left to right) couplings  and corresponding SM background for $pp\to \nu\bar \nu \gamma$ process at HE-LHC.  \label{ptmet_HELHC}}
\end{figure}

Since we focus on $pp\to \nu\nu\gamma$ process to search for sensitivity to the dimension-8 anomalous $Z\gamma\gamma$ and $ZZ\gamma$ couplings, at least one photon is required with non-zero missing energy transverse (MET) in the final state. One can expect to find distinctive properties of this process in the kinematic distributions of leading photon and MET. The $p_T^{\gamma}$ distribution of leading photon (upper) and MET distribution (lower) for signal $C_{\widetilde BW}/\Lambda^4$,$C_{BB}/\Lambda^4$, $C_{WW}/\Lambda^4$  and $C_{BW}/\Lambda^4$ (left to right) couplings  and corresponding SM background for $pp\to \nu\bar \nu \gamma$ process at HL-LHC and HE-LHC are given in Fig.~\ref{ptmet_HLLHC} and Fig.~\ref{ptmet_HELHC}, respectively.  

As it can be seen from Fig.\ref{ptmet_HLLHC} and Fig. \ref{ptmet_HELHC}, the deviation of signal from the SM background for all couplings appears to be around 400 GeV for both $p_T^{\gamma}$ and MET distributions. Therefore, we impose the following cuts  $p_T^{\gamma}$ > 400 GeV,  MET > 400 GeV and $|\eta^{\gamma}|$< 2.5 for further analysis. 

\section{Results of the Analysis}
\begin{figure}[htb!]
\includegraphics[scale=0.6]{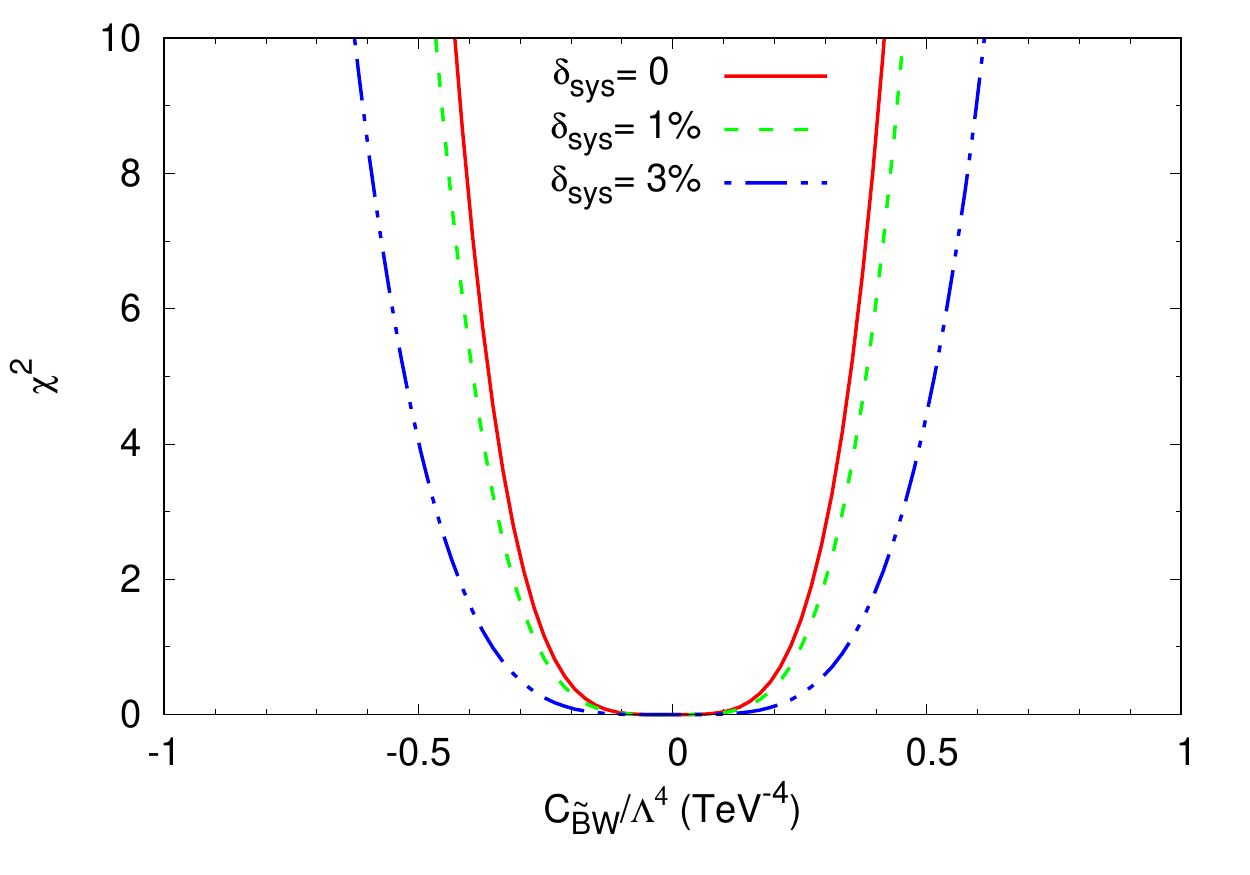} 
\includegraphics[scale=0.6]{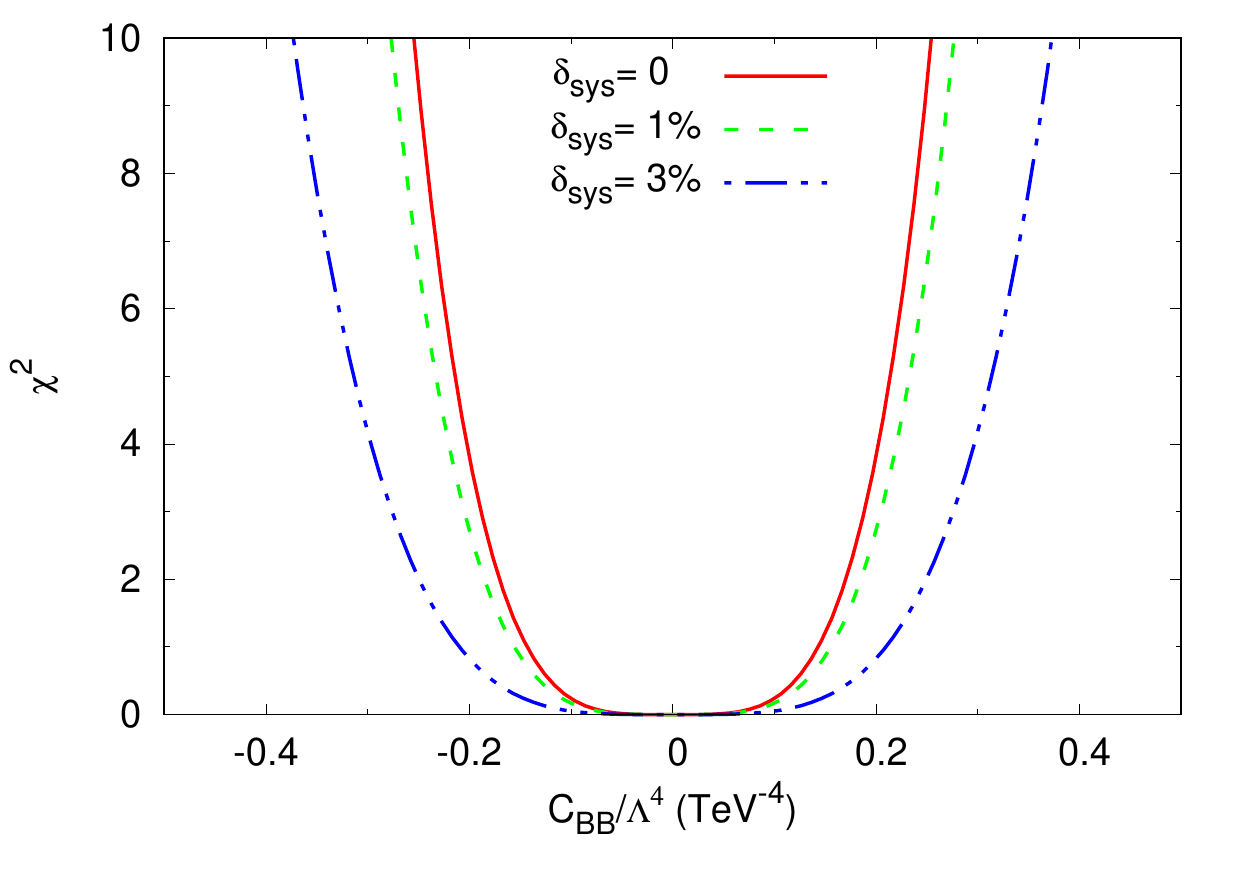} \\
\includegraphics[scale=0.6]{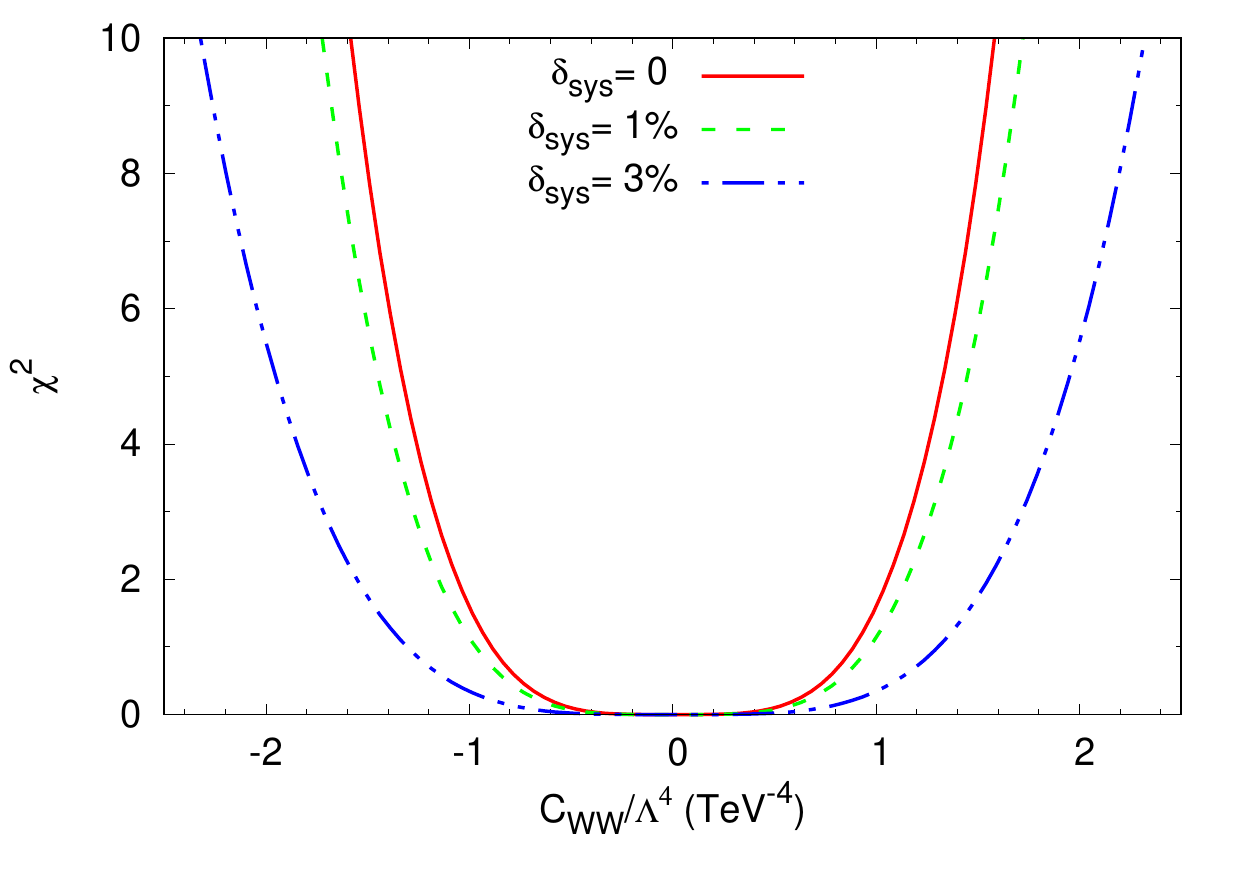} 
\includegraphics[scale=0.6]{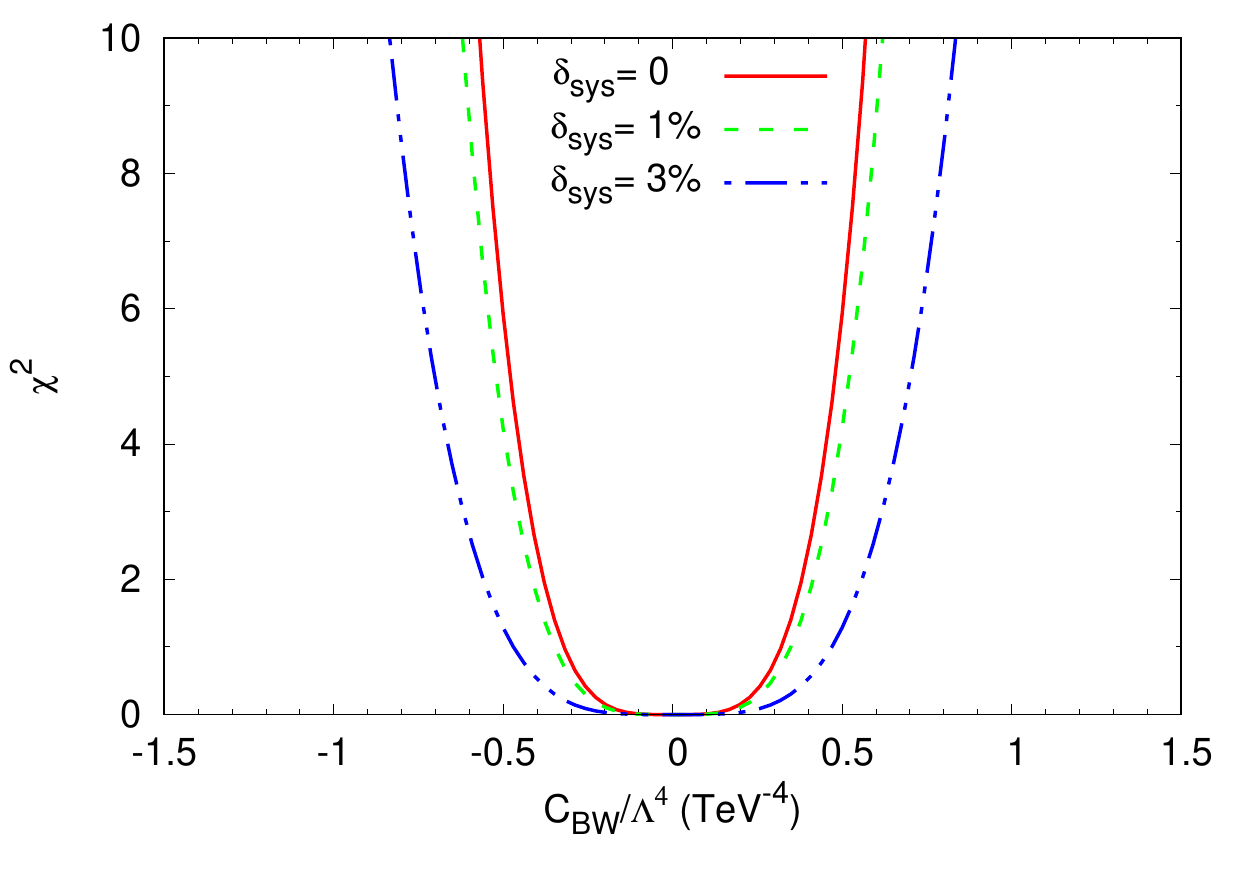}   
\caption{ Obtained $\chi^2$ as a functions of $C_{\widetilde B W}/\Lambda^4$,$C_{B B}/\Lambda^4$ $C_{W W}/\Lambda^4$  and $C_{B W}/\Lambda^4$  couplings at HL-LHC with an integrated luminosity of 3 ab$^{-1}$ including without and with systematic errors (1\% and 3\%). The limits are each derived with all other coefficients set to zero.  \label{chiHLLHC}}
\end{figure}

\begin{figure}[htb!]
\includegraphics[scale=0.6]{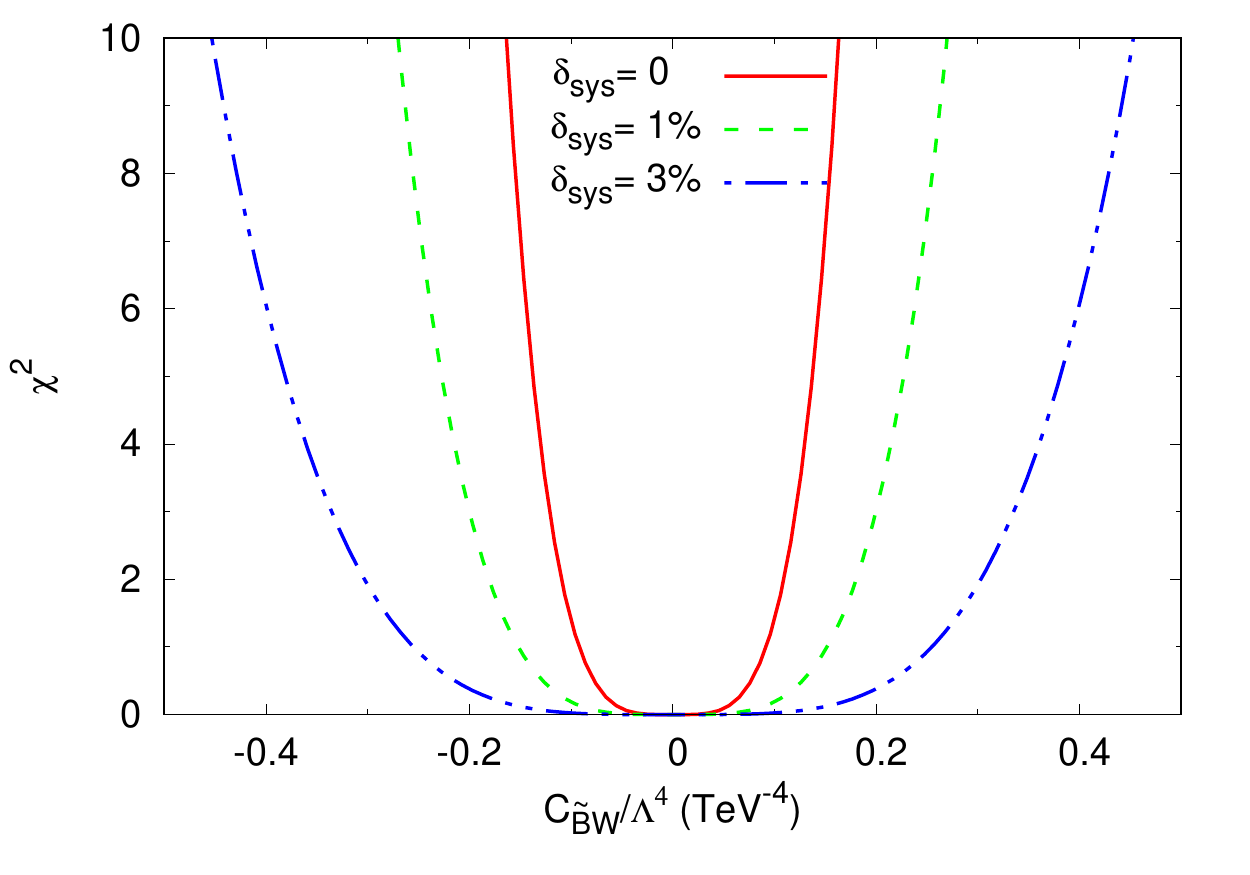} 
\includegraphics[scale=0.6]{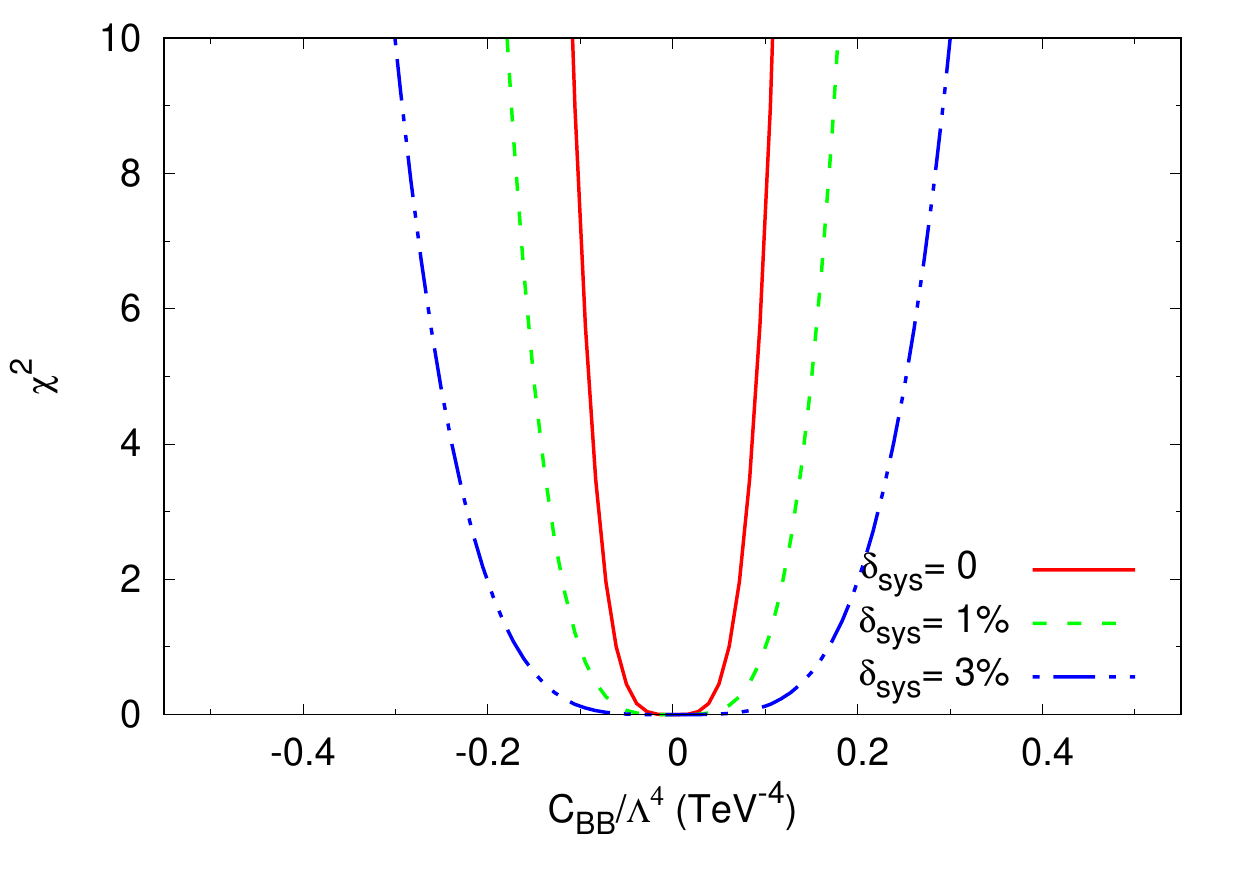} \\
\includegraphics[scale=0.6]{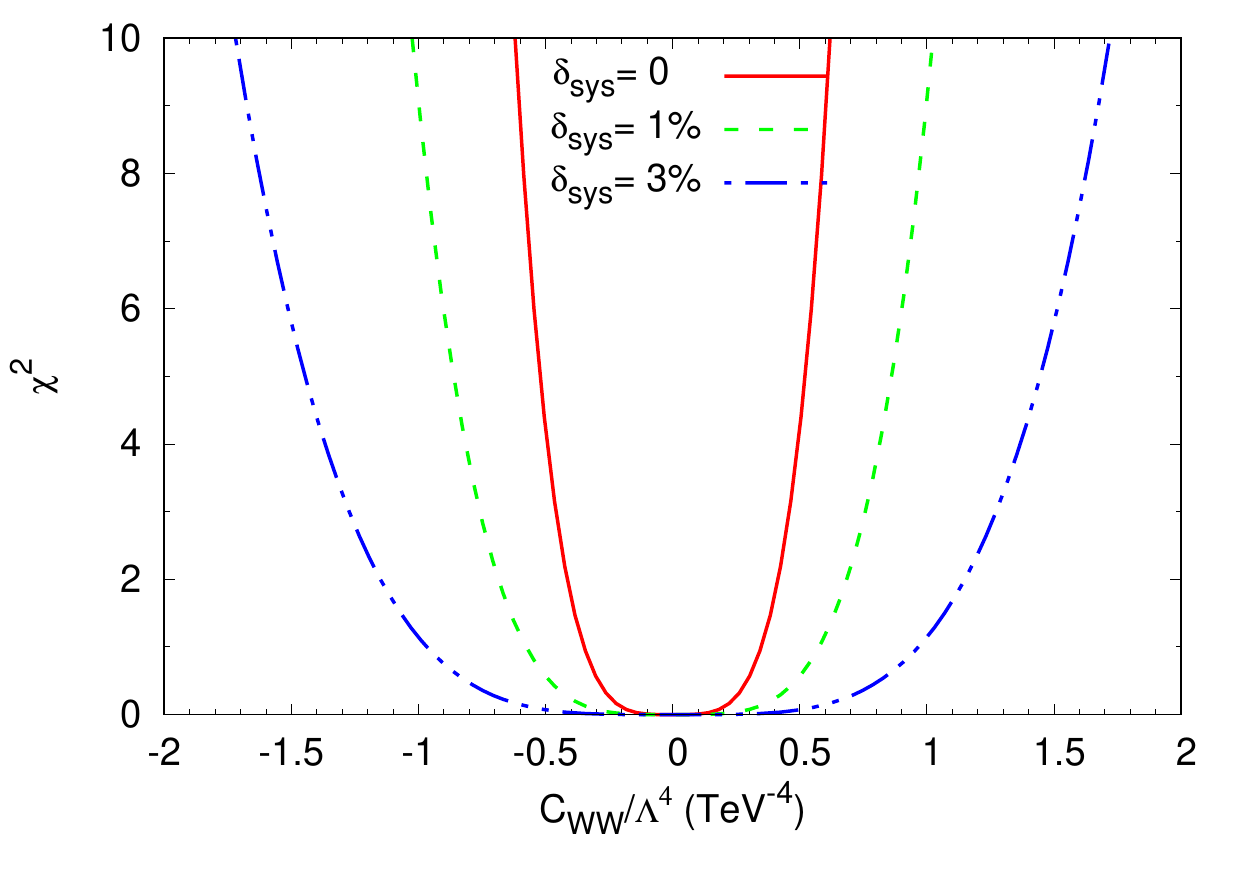} 
\includegraphics[scale=0.6]{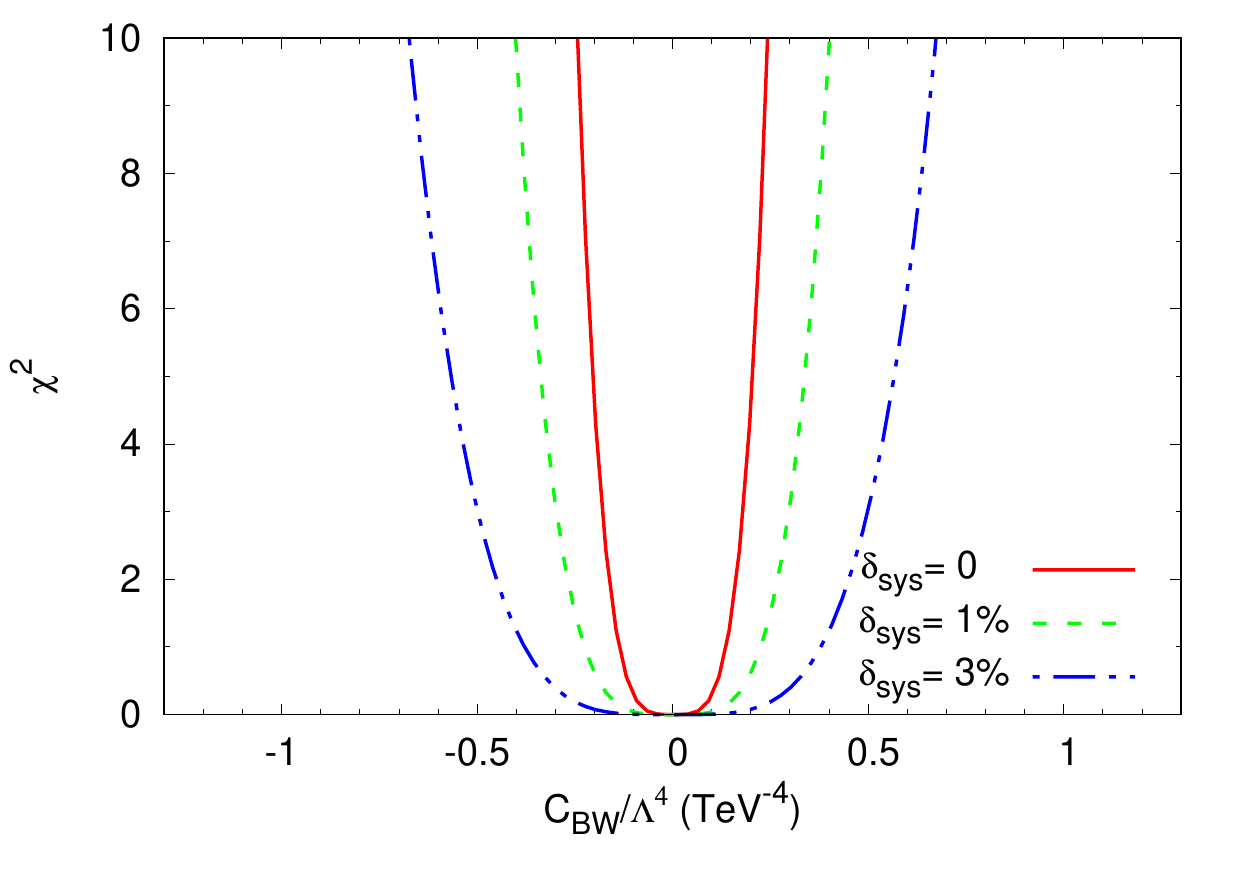}   
\caption{ Obtained $\chi^2$ as a functions of $C_{\widetilde B W}/\Lambda^4$,$C_{B B}/\Lambda^4$ $C_{W W}/\Lambda^4$  and $C_{B W}/\Lambda^4$  couplings at HE-LHC with an integrated luminosity of 15 ab$^{-1}$ including without and with systematic errors (1\% and 3\%). The limits are each derived with all other coefficients set to zero.   \label{chiHELHC}}
\end{figure}

\begin{figure}[htb!]
\includegraphics[scale=0.8]{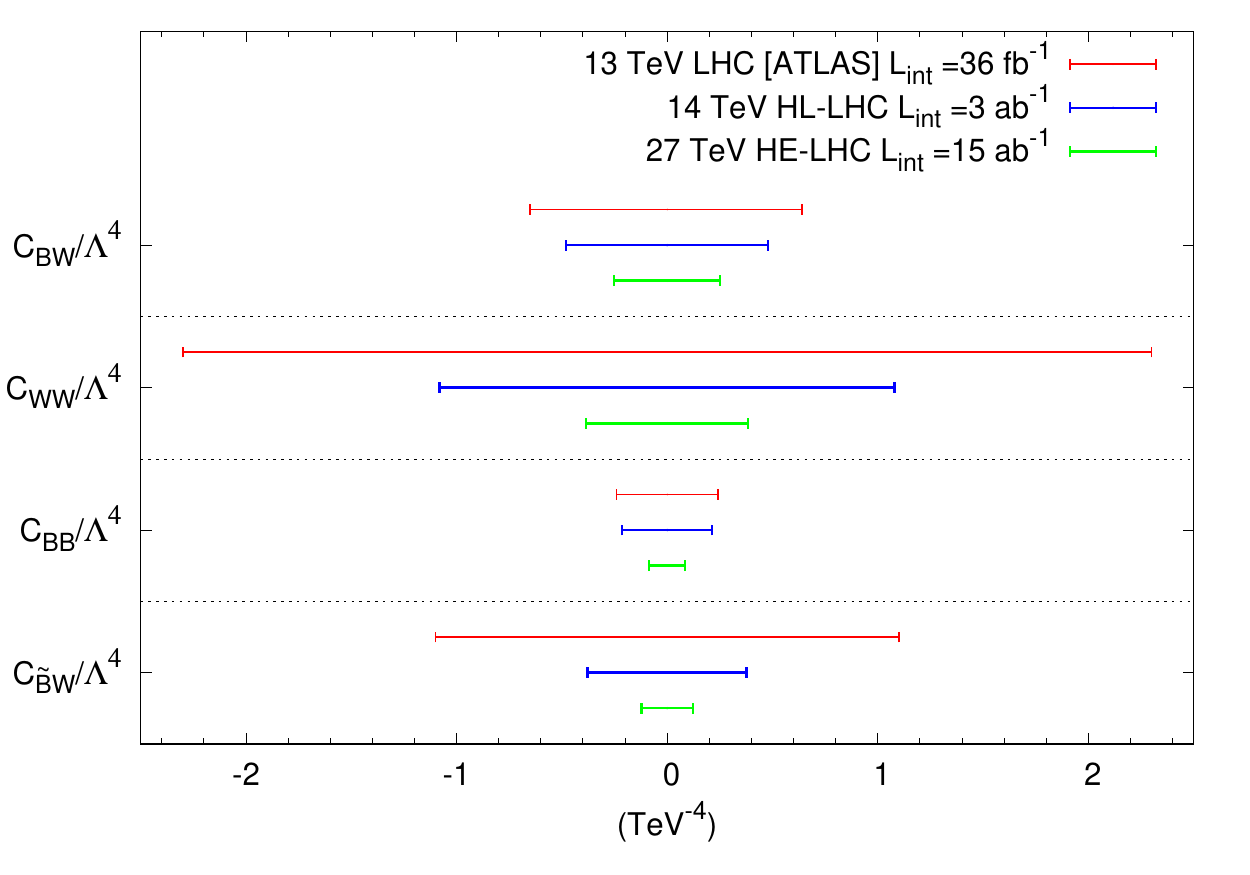}   
\caption{ Comparison of one-dimensional 95\% C.L. intervals in plane for $C_{\widetilde B W}/\Lambda^4$,$C_{B B}/\Lambda^4$ $C_{W W}/\Lambda^4$  and $C_{B W}/\Lambda^4$ without systematic errors at HL-LHC, HE-LHC and LHC  \label{limits}}
\end{figure}

In order to obtain sensitivity of $C_{\widetilde BW}/\Lambda^4$,$C_{BB}/\Lambda^4$, $C_{WW}/\Lambda^4$  and $C_{BW}/\Lambda^4$ couplings via the $pp\to \nu\bar \nu \gamma$ process, we use $\chi^{2}$ method. The $\chi^{2}$ function with and without a systematic error is defined as follows
\begin{eqnarray}
\chi^{2} =\sum_i^{n_{bins}}\left(\frac{N_{i}^{NP}-N_{i}^{B}}{N_{i}^{B}\Delta_i}\right)^{2}
\end{eqnarray}
where $N_i^{NP}$ is the total number of events in the existence of effective couplings, $N_i^B$ is total number of events of the corresponding SM backgrounds in $i$th bin of the $p_T^{\gamma}$ distributions, $\Delta_i=\sqrt{\delta_{sys}^2+\frac{1}{N_i^B}}$ is the combined systematic ($\delta_{sys}$) and statistical errors in each bin. Fig. \ref{chiHLLHC} and Fig. \ref{chiHELHC} show obtained $\chi^2$ as a functions of $C_{\widetilde B W}/\Lambda^4$,$C_{BB}/\Lambda^4$, $C_{WW}/\Lambda^4$  and $C_{BW}/\Lambda^4$  couplings at HL-LHC with an integrated luminosity of 3 ab$^{-1}$ and HE-LHC with an integrated luminosity of 15 ab$^{-1}$ including without and with systematic errors (1\% and 3\%), respectively. Our obtained 95\% Confidence Level (C.L.) limits without systematic error on aNTG $C_{\widetilde BW}/\Lambda^4$,$C_{BB}/\Lambda^4$ $C_{WW}/\Lambda^4$  and $C_{BW}/\Lambda^4$  couplings for HL-LHC (HE-LHC) are [-0.38;0.38] ([-0.12;0.12]), [-0.21;0.21]([-0.085;0.085]), [-1.08;1.08]([-0.38;0.38]) and [-0.48;0.48]([-0.25;0.25]), respectively. Results compared with current ATLAS limits \cite{ATLAS:2018eke} are summarized in Fig.\ref{limits}. For all couplings we obtained better limits than current experimental limits from LHC. Furthermore, our obtained limits on $C_{\widetilde B W}/\Lambda^4$ and $C_{W W}/\Lambda^4$ couplings at 27 TeV center of mass energy with an integrated luminosity 15 ab$^{-1}$ are one order of magnitude better. Even including 3 \% systematic error, our results are better for HE-LHC and comparable for HL-LHC.  One can also perform the sensitivity of the dimension-8 couplings using tagged protons at hadron-hadron colliders as suggested in Ref. \cite{Chapon:2009hh,Senol:2013ym,Sahin:2012mz}.

The new physics characteristic scale $\Lambda$ could be related to the coefficients of dimension-eight operators. Assuming that the underlying theory is strongly coupled, an upper bound can be put on the new physics scale $\Lambda$. For couplings $O(1)$, we find $\Lambda < \sqrt{2\pi v \sqrt{s}} \sim 17.5$ TeV. This upper bound is not violated in this analysis since we have $p_T^{\gamma}<  800 $  GeV for the kinematical range of distributions related to the photon in the final state.

\section{Conclusions}
In this study, we focused on the $pp\to \nu\nu\gamma$ process to search for effects of the dimension-8 anomalous $Z\gamma\gamma$ and $ZZ\gamma$ operators at proposed future post LHC experiments, HL-LHC and HE-LHC. Due to importance of pile-up effects, we included detector effects with the Delphes card which is prepared for upgrade study of LHC with 140 pile-up. It is observed that transverse momentum and pseudo-rapidity of the leading photon and MET are crucial to distinguish signal and background in the cut-based analysis. Using  the transverse momentum of the leading photon of the signal and background processes, the sensitivities of $C_{\widetilde B W}/\Lambda^4$,$C_{B B}/\Lambda^4$ $C_{W W}/\Lambda^4$  and $C_{B W}/\Lambda^4$ couplings were obtained by applying $\chi^{2}$ approach for 3 ab$^{-1}$  and 15 ab$^{-1}$ integrated luminosity of HL-LHC and HE-LHC, respectively. We compared our results with the the ATLAS collaboration limits obtained from the production of Z boson in association with a high energy photon at $\sqrt s$ = 13 TeV with an integrated luminosity of 36.1 fb$^{-1}$  \cite{ATLAS:2018eke} . We conclude that having high luminosity and high energy at post LHC experiments results in better sensitivities to all dimension-8 couplings with and without systematic errors.
 
\begin{acknowledgments}
This work partially supported by the Turkish Atomic Energy Authority (TAEK) under the grant No. 2013TAEKCERN-A5.H2.P1.01-24.
\end{acknowledgments}

\end{document}